\documentclass[reprint,amsfonts,amsmath,amssymb,aps,superscriptaddress]{revtex4-2}
\pdfoutput=1
\usepackage{graphicx}
\usepackage[colorlinks=true,linkcolor=black,anchorcolor=black,citecolor=black,filecolor=black,menucolor=black,runcolor=black,urlcolor=black]{hyperref}

\begin{document}

\title{Realistic quantum photonic neural networks}

\author{Jacob Ewaniuk}
\email{jacob.ewaniuk@queensu.ca}
\affiliation{Department of Physics, Engineering Physics \& Astronomy, 64 Bader Lane, Queen's University, Kingston, Ontario, Canada K7L 3N6}

\author{Jacques Carolan}
\affiliation{Wolfson Institute for Biomedical Research, University College London, London WC1E 6BT, UK}

\author{Bhavin J. Shastri}
\affiliation{Department of Physics, Engineering Physics \& Astronomy, 64 Bader Lane, Queen's University, Kingston, Ontario, Canada K7L 3N6}
\affiliation{Vector Institute, Toronto, Ontario, Canada, M5G 1M1}

\author{Nir Rotenberg}
\affiliation{Department of Physics, Engineering Physics \& Astronomy, 64 Bader Lane, Queen's University, Kingston, Ontario, Canada K7L 3N6}

\date{\today}

\begin{abstract}
Quantum photonic neural networks are variational photonic circuits that can be trained to implement high-fidelity quantum operations. However, work-to-date has assumed idealized components, including a perfect $\pi$ Kerr nonlinearity. Here, we investigate the limitations of realistic quantum photonic neural networks that suffer from fabrication imperfections leading to photon loss and imperfect routing, and weak nonlinearities, showing that they can learn to overcome most of these errors. Using the example of a Bell-state analyzer, we demonstrate that there is an optimal network size, which balances imperfections versus the ability to compensate for lacking nonlinearities. With a sub-optimal $\pi/10$ effective Kerr nonlinearity, we show that a network fabricated with current state-of-the-art processes can achieve an unconditional fidelity of 0.891, that increases to 0.999999 if it is possible to precondition success on the detection of a photon in each logical photonic qubit. Our results provide a guide to the construction of viable, brain-inspired quantum photonic devices for emerging quantum technologies.
\end{abstract}

\maketitle

\section{Introduction} \label{sec:intro}
Quantum neural networks, brain-inspired quantum circuits, harness artificial intelligence to enhance quantum information processing.  When driven with light, quantum photonic neural networks (QPNNs) leverage the strengths of mature photonic platforms \cite{Killoran:19}, including multiplexing, low latency, and ultra-low operational powers already being exploited by conventional neural networks \cite{Shastri:21} and linear-optical quantum processors \cite{Wang:18}. This allows QPNNs to perform quantum state tomography \cite{Torlai:18}, act as quantum simulators \cite{Aspuru-Guzik:12, Sparrow:18}, process \cite{Steinbrecher:19} or reduce the noise \cite{Bondarenko:20} of quantum states, or speed up tasks normally carried out by classical neural networks, such as image recognition \cite{Parthasarathy:21} and natural language processing \cite{DiSipio:22}.

\begin{figure*}[htb!]
\centering
\includegraphics[width=\textwidth]{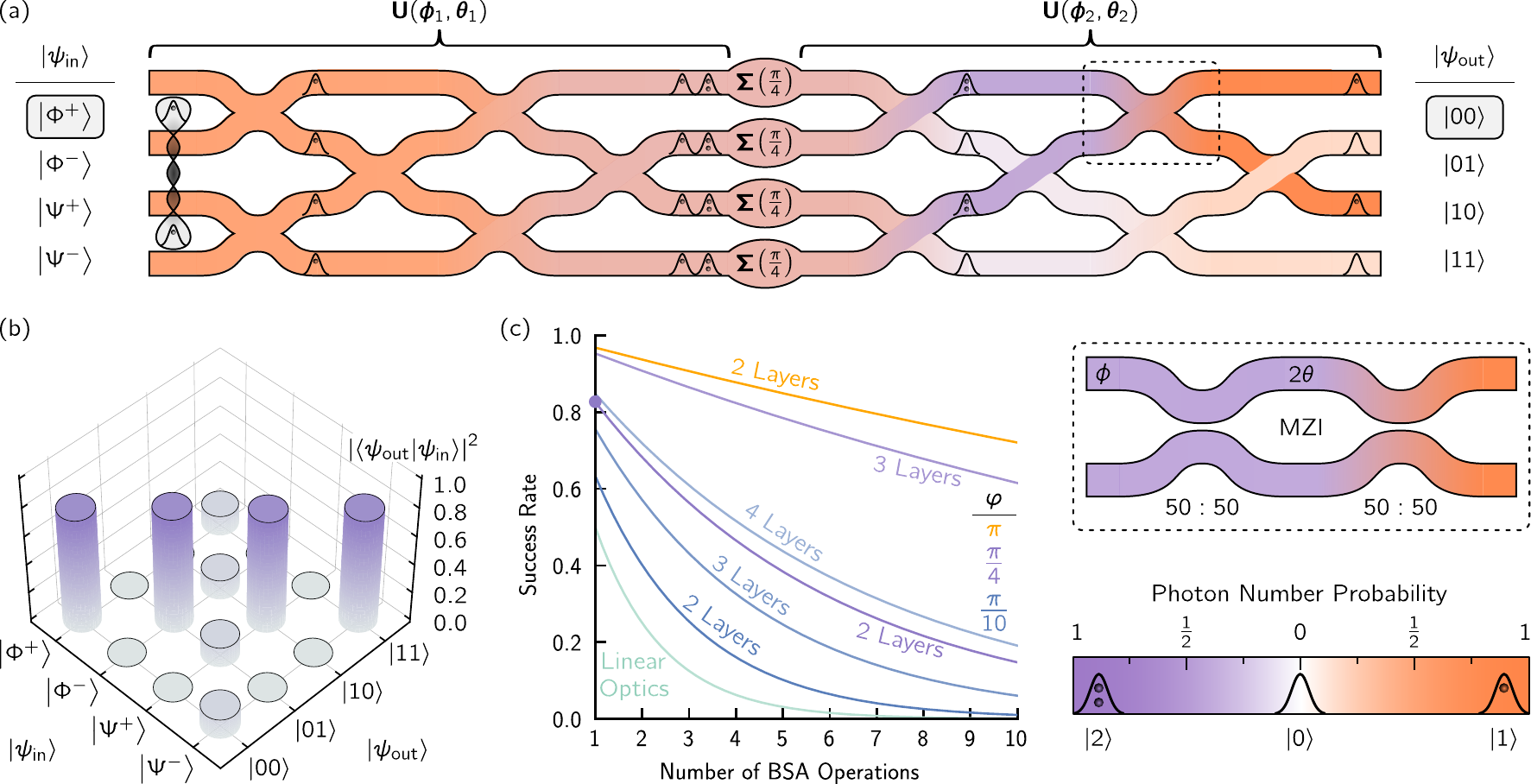}
\caption{A realistic QPNN-based BSA. (a) An exemplary two-layer QPNN consisting of meshes $\left(\mathbf{U}\right)$ of parameterized Mach-Zehnder interferometers (inset) separated by single-site nonlinearities $\left(\boldsymbol{\Sigma}\right)$. The network features two dual-rail encoded qubits, one where a single photon occupies the upper two spatial modes, the other in the lower two modes. Here, realistic losses (0.3 dB/cm) and errors in the $50:50$ directional couplers (5.08\%), as well as a weak $\pi/4$ nonlinear phase shift are assumed. The network was trained to act as a BSA according to the truth table shown, with a resultant unconditional fidelity of 0.825. As an example, the network is colored to portray the propagation of the photons through the network when the $\left|\Phi^+\right\rangle$ Bell-state is incident. The colors represent the probabilities that there are zero, one, or two photons in each spatial mode at each part of the network (colorbar), showing the evolution of the state as it propagates through the circuit. For this example, there is an 82.5\% chance of measuring the correct $\left|00\right\rangle$ target state. (b) Probabilities of measuring a state $\left|\psi_\mathrm{out}\right\rangle$ when a state $\left|\psi_\mathrm{in}\right\rangle$ is fed into the network shown in (a). (c) Comparison between the success rates of ideal linear-optical and realistic QPNN-based BSAs when up to ten are operated in series. The linear-optical BSA has a maximal unconditional fidelity of 0.5 \cite{Calsamiglia:01} and is compared to realistic QPNNs with varying amounts of layers and effective nonlinear phase shifts ($\varphi$), as explained in the main text. The purple marker highlights the success rate for the network shown in (a).}
\label{fig:overview}
\end{figure*}

An example of a QPNN circuit, a two-layer network trained to act as a Bell-state analyzer (BSA), is shown in Fig.~\ref{fig:overview}a.
Here, the connectivity and activation function for the network are provided by linear, rectangular interferometer meshes $\left(\mathbf{U}\right)$ \cite{Clements:16} and single-site optical nonlinearities $\left(\boldsymbol{\Sigma}\right)$, respectively. A BSA can distinguish between, or create, all four highly entangled Bell-states, and the addition of this nonlinearity ideally increases the success probability of the circuit to unity \cite{Steinbrecher:19} from 0.5 as possible solely with linear optics in the absence of ancillary photons \cite{Calsamiglia:01}. The operation performed by the QPNN in this example is therefore crucial to entanglement swapping \cite{Zukowski:93} and hence provides a route toward a deterministic quantum repeater node \cite{Azuma:15}, a vital component of a future quantum internet \cite{Kimble:08}.

Quantum photonic circuits are not ideal, and here we report on the performance of realistic, imperfect QPNNs. Specifically, we consider how propagation losses and imperfect optical nonlinearities affect the fidelity of the QPNN, using the example of a BSA to benchmark our results. As shown in Figs.~\ref{fig:overview}b and c, we find that even realistic networks with weak nonlinearities can vastly outperform those based on (ideal) linear optics. In Fig.~\ref{fig:overview}c, we observe that using state-of-the-art waveguide fabrication (as described in Sec.~\ref{sec:arch}) and a perfect $\left(\pi\right)$ two-photon nonlinearity, 10 BSA nodes made from 2-layer QPNNs can be applied in series with a success rate of $72\%$. Surprisingly, this rate is only decreased to $61\%$, for a much weaker $\left(\pi/4\right)$ two-photon nonlinearity if a third (lossy) layer is added. Moreover, if each operation is conditioned on the measurement of two photons, the conditional success rate of 10 nodes becomes $99.99999\%$ and $99.9\%$ for $\pi$ and $\pi/4$, respectively. In what follows, we unravel the dependence on loss, effective nonlinearity, and network size, providing a methodology for the design of optimal QPNNs with realistic components.

\section{Network Architecture \& Nonidealities} \label{sec:arch}
The architecture of a realistic QPNN is the same as that of an ideal network, and is thus designed to operate on dual-rail encoded photonic qubits \cite{Steinbrecher:19}. Each layer consists of a mesh of tunable Mach-Zehnder interferometers (MZIs) with two controllable phase shifters $\left(\phi,\theta\right)$, as shown in the inset to Fig.~\ref{fig:overview}a. The interferometer mesh can be programmed to perform any arbitrary linear unitary transformation $\mathbf{U}\left(\boldsymbol{\phi},\boldsymbol{\theta}\right)$ on the spatial modes of the photons \cite{Clements:16}. Single-site nonlinearities, of strength $\varphi$, are placed between consecutive layers. In the Supplementary Information S1, we provide further details on the construction of the system transfer function.

The components of linear photonic networks are not perfect, and various techniques have been developed to mitigate the effects of these imperfections. Specifically, both imperfect splitting ratios of the directional couplers (DCs) that form the MZIs and imperfectly calibrated phase shifters lead to errors that can be mitigated by optimizing the circuits after fabrication \cite{Miller:15, Mower:15, Hamerly:21}. In contrast, we account for these errors and those due to imbalanced photon loss or imperfect nonlinearities by training the variational parameters $\left\{\boldsymbol{\phi}_{i},\boldsymbol{\theta}_{i}\right\}$ \emph{in situ}, as would be done on-chip, post-fabrication.

We model a realistic linear mesh by allowing each element to suffer from slightly different imperfections, resulting in unbalanced, photon-path-dependent errors. We define the transmittance of each DC as $t$, randomly selected from a normal distribution with a mean of 50\% and standard deviation of 5.08\%, matching experimental results of a broadband DC fabricated for silicon-on-insulator (SOI) platforms \cite{Lu:15}. Likewise, propagation losses, where photons are scattered out of the circuit due to, for example, surface roughness, or are absorbed by the waveguides, are parameterized by
\begin{equation} \label{eq:alpha}
    \alpha=1-10^{-\frac{\alpha_{\mathrm{WG}}\ell}{10}},
\end{equation}
for an element of length $\ell$ and propagation losses per unit length $\alpha_{\mathrm{WG}}$. These losses depend on the platform upon which the photonic circuit is constructed \cite{Ding:12, Worhoff:15, Melchiorri:05, Cai:15, Nezhad:11, DAgostino:15, Cardenas:09}, which additionally determines the size of each photonic element. In our analysis, we select $\alpha_\mathrm{WG}$ from a normal distribution with a standard deviation of $6.67\%$ of the mean, as is the case for current state-of-the-art photonic circuits built on SOI, which suffer from $\alpha_\mathrm{WG}=0.3 \pm 0.02$ dB/cm at 1550 nm \cite{Cardenas:09}. More information on the inclusion of fabrication imperfections can be found in Sec.~\ref{sec:meth} and the Supplementary Information S1.

In our architecture, as in previous realizations \cite{Steinbrecher:19}, a Kerr nonlinearity, resolved in the Fock basis as,
\begin{equation} \label{eq:kerr}
    \boldsymbol{\Sigma}\left(\varphi\right)=\sum_{n}\exp\left[in(n-1)\frac{\varphi}{2}\right]\left|n\right\rangle \left\langle n\right|, 
\end{equation}
is assumed, wherein the ideal case $\varphi = \pi$ such that a single photon passing through will experience no phase change while two photons will undergo a $\pi$ phase change. To date, a $\pi$ Kerr-nonlinearity has yet to be observed at the single-photon level; however, this efficiency has been reached by other nonlinearities, such as those based on electromagnetically induced transparency \cite{Zuo:19} or the saturation \cite{Guo:21} of atoms. While these enable neural networks capable of quantum state tomography \cite{Zuo:22} or image recognition \cite{Ryou:21} respectively, neither are scalable nor compatible with most quantum information processing applications since they lead to photon loss. It is, however, likely that in the near future, efficiencies approaching $\pi$ will be demonstrated, either through the coherent chiral scattering of photons from single quantum emitters \cite{Lodahl:17}, or using integrated nanophotonic cavities designed to address this need \cite{Choi:17, Heuck:20-1, Heuck:20-2}. Hence, we consider single-site Kerr nonlinearities but examine the performance of the QPNN also in the realistic scenario where $\varphi \lesssim \pi$.

In sum, the total transfer function for an $L$-layer QPNN is given by,
\begin{equation} \label{eq:S}
    \mathbf{S} = \mathbf{U}\left(\boldsymbol{\phi}_L, \boldsymbol{\theta}_L\right) \cdot \prod_{i=1}^{L-1}\boldsymbol{\Sigma}\left(\varphi\right)\cdot\mathbf{U}\left(\boldsymbol{\phi}_i,\boldsymbol{\theta}_i\right).
\end{equation}
This transfer function will act on any input state to produce an actual output state $\left|\psi_{\mathrm{out,act}}^{(i)}\right\rangle = \mathbf{S}\left|\psi_{\mathrm{in}}^{(i)}\right\rangle$, which is compared to the ideal output state $\left|\psi_{\mathrm{out}}^{(i)}\right\rangle$ to determine the unconditional fidelity for that input-output pair,
\begin{equation} \label{eq:Func_i}
    \mathcal{F}^{(\mathrm{unc})}_i=\left|\left\langle \psi_{\mathrm{out}}^{(i)}\right|\mathbf{S}\left|\psi_{\mathrm{in}}^{(i)}\right\rangle\right|^2.
\end{equation}
The total unconditional fidelity, the chance that the network provides the correct output state for any given input state without preconditions, is then found by averaging over all $K$ input-output pairs according to
\begin{equation} \label{eq:Func}
    \mathcal{F}^{(\mathrm{unc})}=\frac{1}{K}\sum_{i=1}^{K}\mathcal{F}^{(\mathrm{unc})}_i.
\end{equation}
Conversely, we can calculate the unconditional infidelity (i.e. network error) according to $\mathcal{C}^{(\mathrm{unc})} = 1 - \mathcal{F}^{(\mathrm{unc})}$, which we minimize in training the QPNN, using the local gradient-free BOBYQA nonlinear optimization algorithm \cite{BOBYQA}, as available in the NLopt library \cite{NLopt}. Further details on network optimization are provided in Sec.~\ref{sec:meth}.

The success of the network may be conditioned on the detection of photons only in ports which abide by the dual-rail encoding scheme, as in this case we know that a logical output was produced. For the BSA shown in Fig.~\ref{fig:overview}a, this means that a single photon was detected in one of the top two modes and the other in the bottom two modes. We call this the conditional fidelity $\mathcal{F}^{(\mathrm{con})}$, each $i^\mathrm{th}$ term of which is related to the unconditional fidelity by the probability that the network operation results in a computational basis state (that is, no photons are lost and each logical photonic qubit contains a single photon at the output) $\mathcal{P}^{(cb)}$ by,
\begin{equation} \label{eq:probs}
    \mathcal{F}_i^{(\mathrm{unc})} = \mathcal{F}_i^{(\mathrm{con})}\mathcal{P}_i^{(\mathrm{cb})}.
\end{equation}
In Sec.~\ref{sec:meth}, we provide the expressions used to calculate these conditional measures. For a network operation that requires detection, such as the BSA and not, for example, the realization of a quantum logic gate for quantum computation, $\mathcal{P}^{(\mathrm{cb})}$ gives the probability that the network successfully performed its task, while $\mathcal{F}^{(\mathrm{con})}$ provides the quality of the result; in practice, one may optimize on either the conditional or unconditional infidelities, depending on the task under consideration. In the following, we train solely on the unconditional, however, we provide results from optimizing the conditional infidelity in the Supplementary Information S2.

\section{Correcting Imperfect Linear Interferometer Meshes} \label{sec:ilim}
\begin{figure*}[ht!]
\centering
\includegraphics[width=\textwidth]{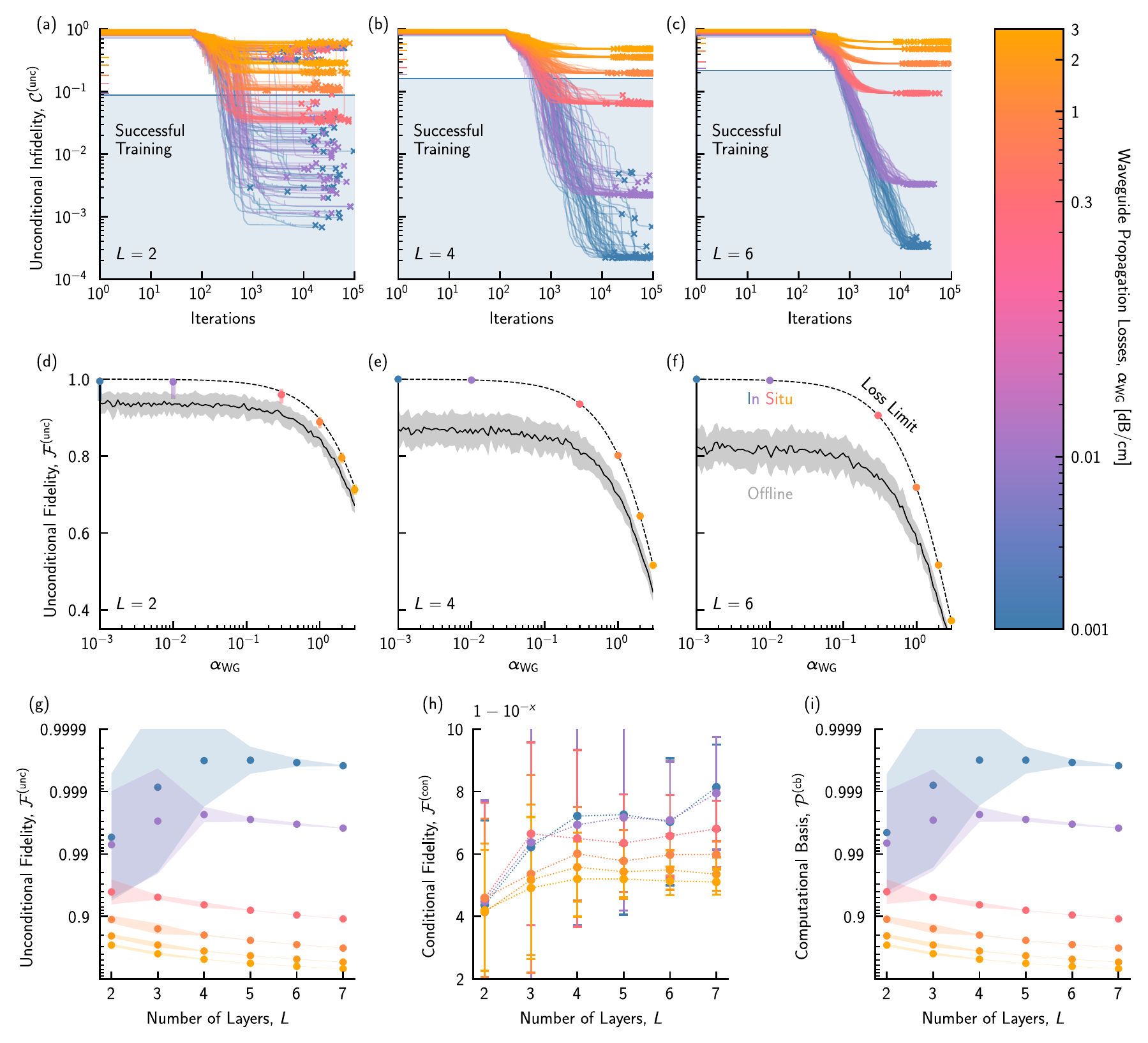}
\caption{Performance of a QPNN-based BSA suffering from fabrication imperfections. The unconditional infidelity $\mathcal{C}^{(\mathrm{unc})}$ of (a) 2, (b) 4, and (c) 6-layer networks are shown as a function of the training iteration for increasingly lossy networks. In each pane, the results of 50 optimization trials are displayed, with clear plateaus visible in $\mathcal{C}^{(\mathrm{unc})}$ that increase with the losses. In each case, only trials that result in infidelity at or below those achieved by offline training (colored ticks in (a)-(c), shaded regions in (d)-(f)) are considered successful (shaded blue region shows an example for 0.001 dB/cm). The unconditional fidelity $\mathcal{F}^{(\mathrm{unc})}$ of (d) 2, (e) 4, and (f) 6-layer networks are plotted with respect to the average losses $\alpha_\mathrm{WG}$, with colored symbols (shaded regions) corresponding to the mean ($95\%$ confidence interval) of a logarithmic normal distribution fitted to the successful trials of (a)-(c) (see the Supplementary Information S3 for more details). These points are seen to lie on the (dashed) loss limit curve, where the performance of the network is only limited by uniform photon loss (assumes perfect DCs; see Sec.~\ref{sec:meth} for additional details), in contrast to networks that are trained offline (solid black curves and shaded grey regions), demonstrating the ability of QPNNs to learn to overcome imperfections. (g) Unconditional fidelity $\mathcal{F}^{(\mathrm{unc})}$, (h) conditional fidelity $\mathcal{F}^{(\mathrm{con})}$, and (i) computational basis probability $\mathcal{P}^{(\mathrm{cb})}$, as a function of $L$ for the QPNNs trained \emph{in situ}, where the mean (symbols) and 95\% confidence intervals (shaded regions in (g), (i), error bars in (h)) are determined via the same method as (d)-(f).}
\label{fig:loss}
\end{figure*}

We begin to consider the effects of imperfections on QPNNs by holding the nonlinearity at the ideal value ($\varphi = \pi$) but introducing DC splitting ratio variations and photon loss as described above. The resultant unconditional infidelity of a BSA for 2 to 6-layer QPNNs with losses ranging from 0.001 to 3 dB/cm, as a function of training iterations, is shown in Figs.~\ref{fig:loss}a-c.
Each case is repeated 50 times, resulting in plateaus of $\mathcal{C}^{(\mathrm{unc})}$ that increase in value for increasing losses, as expected. Interestingly, for 2-layer BSAs, we observe a large spread in the final $\mathcal{C}^{(\mathrm{unc})}$, particularly for low-loss networks (c.f. blue and purple curves in Fig.~\ref{fig:loss}a), indicating that the final performance of the QPNN is largely dictated by imbalance due to imperfect DCs. Adding more layers to the network, as in Fig.~\ref{fig:loss}b and c, reduces this spread, showing that larger QPNNs may learn to correct for these errors and more often reach optimal performance.

This is reflected in Figs.~\ref{fig:loss}d-f, which show the unconditional fidelity $\mathcal{F}^{(\mathrm{unc})}$ as a function of waveguide loss for different sized networks. Here, we compare the \emph{in situ} trained QPNNs of Figs.~\ref{fig:loss}a-c, denoted by the symbols, to the case where the network is trained offline. Offline training means that a perfect network was trained, then losses and DC errors were subsequently added to the solution. This was repeated 50 times for each $\alpha_\mathrm{WG}$, selecting different random imperfections at each repetition, with the mean given by the solid-black curve and standard deviation by the grey region. When trained \emph{in situ}, the QPNN learns to overcome these imperfections as is seen by the convergence toward loss-limited performance (see Sec.~\ref{sec:meth} for more information on network training and the loss limit). This is more apparent for larger networks, where the fidelity of those trained offline significantly reduces due to increased losses and DC errors, while those trained \emph{in situ} maintain and even increase $\mathcal{F}^{(\mathrm{unc})}$.

The balance between fabrication imperfections and network size, as a function of losses, is summarized in Fig.~\ref{fig:loss}g. Here, we observe that for state-of-the-art losses $\left(\alpha_\mathrm{WG} = 0.3\text{ dB}/\text{cm}\right)$ or worse, the unconditional fidelity decreases as expected when more layers are added to the network. When $\alpha_\mathrm{WG} = 0.3\text{ dB}/\text{cm}$,  $\mathcal{F}^{(\mathrm{unc})} \geq 0.905$ (0.904), where the bracketed result is the lower bound of the 95\% confidence interval, even for a 6-layer QPNN, demonstrating that high-efficiency performance is possible on realistic state-of-the-art systems. Conversely, a more complex evolution is seen in Fig.~\ref{fig:loss}g for 0.01 dB/cm losses or less, where there exists an optimal network size other than 2 layers. Losses at 0.01 dB/cm  are similar to those of the silicon nitride platform for 1550 nm, reported as low as 0.007 dB/cm \cite{Bauters:11}, but more typically near 0.01 dB/cm \cite{Shaw:05, Blumenthal:18, Liu:21}. In this low-loss case $\left(\alpha_\mathrm{WG} = 0.01\text{ dB}/\text{cm}\right)$, $\mathcal{F}^{(\mathrm{unc})}$ first increases from 0.993 (0.949) to 0.998 (0.997) by adding two layers to the base size, as the network is better able to account for imperfections. The unconditional fidelity then only slightly decreases to 0.996 (0.996) as the network grows to 7 layers. That is, near-deterministic QPNN-based quantum elements such as BSAs will be realistic in the near-future as platform losses continue to decrease.

The situation is even more promising if the success of the network is preconditioned on detection in the computational basis, as is shown in Figs.~\ref{fig:loss}h and i. Here we present $\mathcal{F}^{(\mathrm{con})}$ and $\mathcal{P}^{(\mathrm{cb})}$ for different sized QPNNs, and for differing $\alpha_\mathrm{WG}$. Even for the extremely lossy networks, where $\alpha_{\mathrm{WG}}= 2\text{ dB}/\text{cm}$, $\mathcal{F}^{(\mathrm{con})}$ remains above 0.9999 (0.9888) for all $L \leq 7$, while for state-of-the-art losses this conditional fidelity does not drop below 0.999999 (0.999784) for $3\leq L\leq 7$, as we observe in Fig.~\ref{fig:loss}h. In fact, as shown in Fig.~\ref{fig:loss}i, it is mainly the rate at which the QPNN produces a logical output that is affected by an increase in network size, showing the potential of even lossy networks if fault-tolerant protocols are used.

\section{Embracing Weak Nonlinear Interactions} \label{sec:wni}
\begin{figure*}[ht!]
\centering
\includegraphics[width=\textwidth]{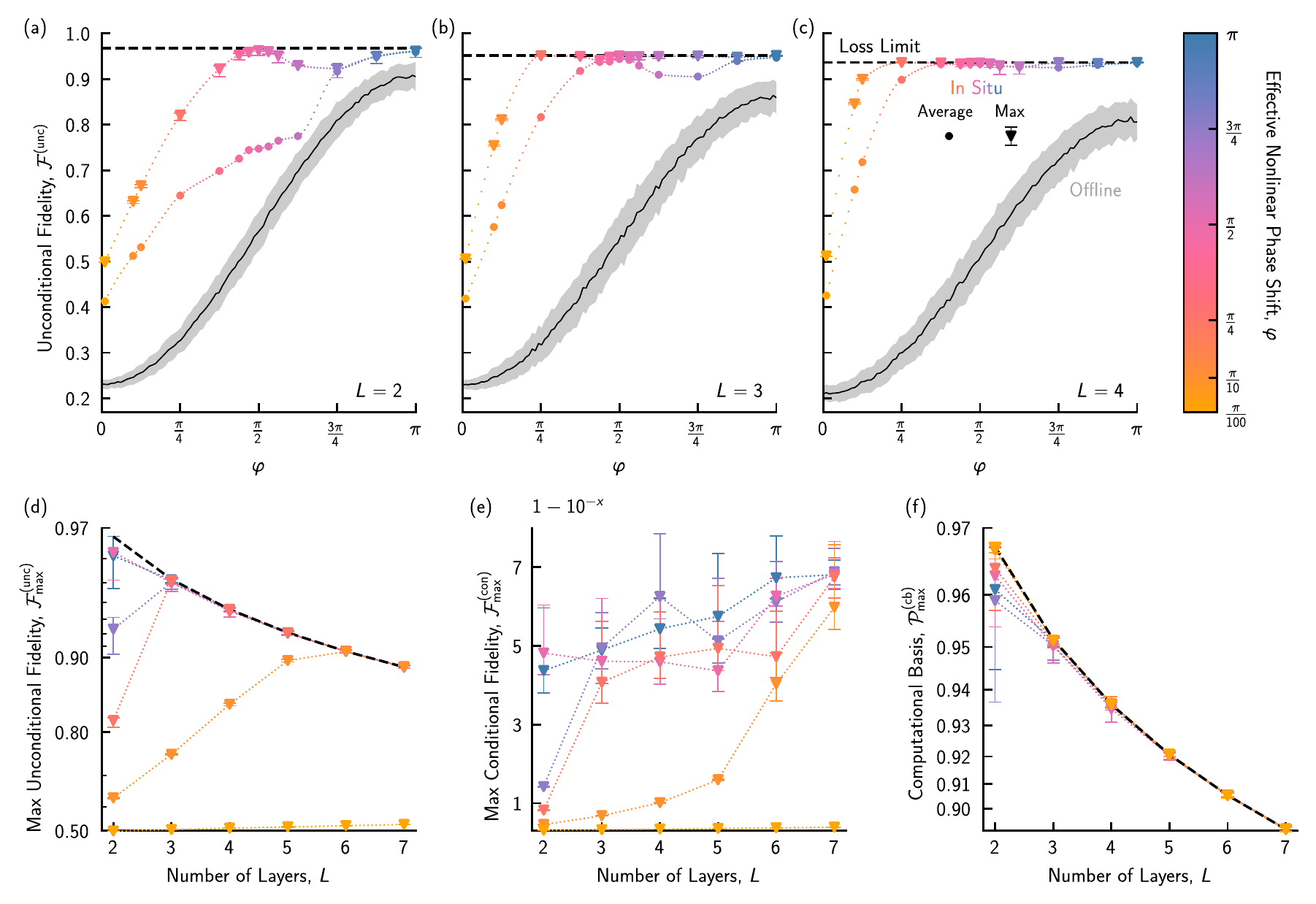}
\caption{Performance of realistic QPNN-based BSAs with sub-optimal $\left(\varphi \lesssim \pi\right)$ nonlinearities and state-of-the-art $\left(\alpha_\mathrm{WG} = 0.3\text{ dB}/\text{cm}\right)$ losses. The unconditional fidelity $\mathcal{F}^{(\mathrm{unc})}$ of (a) 2, (b) 3, and (c) 4-layer networks is shown with respect to the effective nonlinear phase shift $\varphi$, showing both offline (solid black curves, shaded grey regions) and \emph{in situ} (colored symbols) trained networks, and the loss limit (dashed line), as in Fig.~\ref{fig:loss}. \emph{In situ} results include the average of all successfully-trained QPNNs (circles) and the best-case, where triangles (error bars) show the mean (95\% confidence intervals) of a beta distribution fit to the maximal unconditional fidelity plateau (see the Supplementary Information S3, S4 for statistical analysis details and an example of this plateau). The (d) unconditional fidelity $\mathcal{F}^{(\mathrm{unc})}$, (e) conditional fidelity $\mathcal{F}^{(\mathrm{con})}$, and (f) computational basis probability $\mathcal{P}^{(\mathrm{cb})}$ are plotted for each $\varphi$ denoted on the colorbar, for networks of up to 7 layers. All means (triangles) and 95\% confidence intervals (error bars) were determined in the same manner as the best-case \emph{in situ} results of (a)-(c). Connecting dotted lines serve only as a visual aid.}
\label{fig:nl}
\end{figure*}

Having studied the effect of fabrication errors on network performance, we now turn to the consequences of sub-optimal nonlinearities. Assuming state-of-the-art losses $\left(\alpha_{\mathrm{WG}}= 0.3\text{ dB}/\text{cm}\right)$, we vary the effective nonlinear phase shift $\varphi$ from the ideal $\pi$ to $\pi / 100$ and attempt to train QPNNs of different sizes to act as BSAs (see the Supplementary Information S4 for exemplary training traces, c.f. Figs.~\ref{fig:loss}a-c.) For each network size and effective nonlinearity, we attempt to train 200 QPNNs, showing the results in Fig.~\ref{fig:nl}.
Figs.~\ref{fig:nl}a-c depict the highly non-trivial dependence of $\mathcal{F}^{(\mathrm{unc})}$ on the effective nonlinearity $\varphi$. When the QPNN is trained offline, $\mathcal{F}^{(\mathrm{unc})}$ increases monotonically with $\varphi$, as would be the case for a quantum-optical Fredkin gate-based BSA \cite{Milburn:89, Nielsen:11, Gerry:04} (see the Supplementary Information S4 for additional information). Conversely, a QPNN can be trained to account for the weak nonlinearity, in which case it can vastly outperform this expectation. Considering a 2-layer network (Fig.~\ref{fig:nl}a), we observe a strikingly different $\varphi$ dependence when comparing the best-case \emph{in situ} trained networks (triangles; see the Supplementary Information S3 for statistical analysis information) to both the average of all successful \emph{in situ} training cases (circles) and networks trained offline. We observe that networks trained \emph{in situ} can reach the loss limit with sub-optimal nonlinearities, in addition to fabrication imperfections. Specifically, we observe optimal performance of 2-layer QPNNs when $\varphi=\pi/2$ in addition to $\pi$. Moreover, as can be seen in Fig.~\ref{fig:nl}a, near-optimal performance is reached for a domain of $\varphi$ centred at $\pi/2$, providing a pathway to robust QPNN-based BSAs without the need for a perfect Kerr nonlinearity. It must be noted, however, that operating with weaker nonlinearities decreases the probability that the QPNN converges at the loss limit during training, as is shown in the Supplementary Information S4. 

For all $\varphi$, a QPNN trained $\emph{in situ}$ learns how to account for weak nonlinearities and thus approach the loss limit. These capabilities improve as redundancies are added via an increase in network size, as visually evident across Figs.~\ref{fig:nl}a-c and summarized in d. By adding a single additional (lossy) layer, QPNNs were trained to within $1.13\%$ of the unconditional fidelity achieved with the ideal nonlinearity, 0.951 (0.950) at $\varphi = \pi$, and within $1.15\%$ of the loss limit, 0.952, for all $\varphi \geq \pi / 4$. Even networks with nonlinearities as weak as $\varphi = \pi / 10$ approach the loss limit at 6 layers, in contrast to the case of $\pi / 100$ where $\mathcal{F}^{(\mathrm{unc})}$ increases only to 0.528 (0.528) at 7 layers from 0.499 (0.498) at 2 layers, essentially acting as a linear-optical BSA \cite{Calsamiglia:01}. Trainability also improves with increased network size, as it becomes easier for the QPNN to find optimal solutions, such that the average unconditional fidelity achieved during \emph{in situ} training approaches the maximum plateau.

In Figs.~\ref{fig:nl}e and f, the conditional fidelity and computational basis probability are shown as a function of $L \leq 7$, for differing $\varphi$. In contrast to the case where photon losses were varied (c.f. Fig.~\ref{fig:loss}), we observe that the behavior of $\mathcal{F}^{(\mathrm{con})}$ strongly depends on $\varphi$. While QPNNs with $\varphi = \pi$ and $\pi/2$  operate with $\mathcal{F}^{(\mathrm{con})} \geq 0.9999 \left(0.9998\right)$ for all $L \leq 7$, networks with nonlinearities near $\pi/4$ and $3\pi / 4$ require at least 3 layers to reach this level, while at $\pi/10$, 6 layers are needed. For all nonlinearities and network sizes, $\mathcal{P}^{(\mathrm{cb})}$ is within 0.009 (0.031) of loss-limited performance, as seen in Fig.~\ref{fig:nl}f, and as expected for a QPNN suffering from state-of-the-art losses (c.f. Fig.~\ref{fig:loss}d-f and i). Altogether, this demonstrates that for each combination of fabrication imperfections and effective nonlinearity, there exists an optimal network size that maximizes $\mathcal{F}^{(\mathrm{unc})}$. While adding layers will always tend to increase $\mathcal{F}^{(\mathrm{con})}$, a balance must be struck with the exponential decrease in $\mathcal{P}^{(\mathrm{cb})}$. In the Supplementary Information S5, we demonstrate a QPNN trained to generate Greenberger-Horne-Zeilinger states, and show that this remains true beyond the BSA application.

\section{Discussion} \label{sec:disc}
We have shown that high-fidelity operation is possible in realistic quantum photonic neural networks based on non-ideal Kerr nonlinearities. Since propagation through these networks leads to inevitable photon loss, their unconditional fidelity ceiling tends to decrease with increasing size. While this loss limit is unavoidable, these networks are able to learn to manage additional errors from non-uniform losses and directional coupler splitting ratio variations, often demonstrating increased fidelity with the addition of imperfect layers. Crucially, we have shown that weak nonlinearities, which are mere fractions of the ideal, are sufficient for near-optimal network performance. Even as these sub-optimal nonlinearities are realized \cite{Lodahl:17, Choi:17}, the desired phase change will likely be accompanied by wave-packet distortions \cite{Shapiro:06, Gea-Banacloche:10}, and although complex solutions based on dynamically-coupled cavities have been proposed \cite{Heuck:20-1, Heuck:20-2}, it remains an open question if, instead, a QPNN may \emph{learn} to overcome them in much the same way it does fabrication imperfections. Already in the work presented here, QPNNs offer a fascinating view of the intricate balance between loss, imperfect photon routing and weak nonlinearity, which we have unravelled to demonstrate how each combination leads to an optimal network geometry. Understanding and respecting this balance will be important, in the near future, as realistic QPNNs are designed and fabricated. 

It is now clear why QPNNs far outperform linear-optical networks. Even with a weak $\pi/4$ effective nonlinearity, they can learn to surpass the 0.5 unconditional fidelity possible with perfect linear optics \cite{Calsamiglia:01}, achieving $\mathcal{F}^{(\mathrm{unc})} = 0.820$ (0.809) at 2 layers (see Figs.~\ref{fig:overview}b and \ref{fig:nl}a), which grows to 0.951 (0.949) with an additional layer (see Fig.~\ref{fig:nl}b). At 6 layers, loss-limited operation, $\mathcal{F}^{(\mathrm{unc})} = 0.891$ (0.890), can be achieved with nonlinearities as weak as $\pi/10$. Returning to Fig.~\ref{fig:overview}c, which summarizes the success rate of operating $N$ BSAs in series, as would be necessary to connect quantum repeater nodes by entanglement swapping \cite{Zukowski:93, Azuma:15}, the performance benefits offered by QPNNs become more apparent. While 10 consecutive perfect linear-optical BSAs have a success rate of just 0.1\%, 6-layer, $\pi/10$ nonlinearity QPNNs reach 31.5\%, and 3-layer, $\pi/4$ networks achieve 60.5\%.

Preconditioning the success of each QPNN-based BSA on the detection of 2 photons in the correct ports, as would be the case for generating cluster states from fusion gates \cite{Browne:05}, allows the much higher conditional fidelities to be leveraged. While  $\mathcal{F}^{(\mathrm{con})}$ for a perfect linear-optical ten-BSA sequence remains at 1, realistic QPNNs of $\pi / 4$ (3 layers) and $\pi / 10$ (6 layers) nonlinearities reach 0.999 (0.997) and 0.99999 (0.99996), respectively. Given that these conditional fidelities are all near-unity, the rather large variations to $\mathcal{F}^{(\mathrm{unc})}$ seen above can be attributed to the operational rate of the circuits (c.f. Eq.~\ref{eq:probs}), which are $315\times$ improved when the perfect linear-optical BSAs are replaced by even 6-layer, $\pi / 10$ QPNNs. Hence, imperfect QPNNs are likely to play a key role in emerging large-scale quantum technologies.

\section{Methods} \label{sec:meth}
\subsection*{Modeling Fabrication Imperfections}
An ideal MZI, as displayed in the inset to Fig.~\ref{fig:overview}a, can be described by a $2\times 2$ matrix,
\begin{align} \label{eq:idealMZI}
    T^{(\mathrm{ideal})} &= \frac{1}{2}\begin{pmatrix}1&-i\\-i&1\end{pmatrix}\begin{pmatrix}e^{i2\theta}&0\\0&1\end{pmatrix}\begin{pmatrix}1&i\\i&1\end{pmatrix}\begin{pmatrix}e^{i\phi}&0\\0&1\end{pmatrix},\nonumber \\
     &= e^{i\theta}\begin{pmatrix}e^{i\phi}\cos{\theta} & -\sin{\theta} \\ e^{i\phi}\sin{\theta} & \cos{\theta} \end{pmatrix},
\end{align}
as is commonly found in the literature \cite{Clements:16, Mower:15}, up to the arrangement of components specified here. To model a realistic MZI, we include two types of imperfections: photon loss due to propagation and an imperfect splitting ratio of the nominally $50:50$ DCs. Imperfect phase shifter calibration is neglected as a QPNN trained \emph{in situ} would intrinsically learn the phase shifts that account for these errors. A photonic element of length $\ell$ introduces the probability $\alpha$ that a photon is lost via propagation through it. By Eq. \ref{eq:alpha}, $\alpha$ depends on the propagation losses per unit length, $\alpha_\mathrm{WG}$, selected from a normal distribution with a width of 6.67\%, corresponding to the state-of-the-art experimental results for SOI \cite{Cardenas:09}. For each MZI, an individual $\alpha$ is computed, then applied through multiplication by the $2\times 2$ matrix,
\begin{equation} \label{eq:lossMatrix}
    \begin{pmatrix} \sqrt{1-\alpha} & 0 \\ 0 & \sqrt{1-\alpha} \end{pmatrix}.
\end{equation}
In the Supplementary Information S1, further details are given for the inclusion of these non-uniform losses, including the characteristic lengths ($\ell$) of each photonic element, and how the lack of unitarity is dealt with in the simulations. Similarly, each imperfect DC has an individual transmittance $t$ that is taken from a normal distribution centered at 0.5 with a standard deviation of 0.0508, matching experimental results of a broadband DC fabricated for SOI platforms \cite{Lu:15}. For a given $t$, the corresponding $2\times 2$ transformation of the DC is,
\begin{equation} \label{eq:dc}
    \begin{pmatrix} \sqrt{t} & \pm i\sqrt{1-t} \\ \pm i\sqrt{1-t} & \sqrt{t} \end{pmatrix}.
\end{equation}
Altogether, these result in a $2\times 2$ transformation describing a realistic MZI,
\begin{widetext}
\begin{align}
    T^{(\mathrm{real})} &= \begin{pmatrix}\sqrt{1-\alpha}&0\\0&\sqrt{1-\alpha}\end{pmatrix}\begin{pmatrix}\sqrt{t_2}&-i\sqrt{1-t_2}\\-i\sqrt{1-t_2}&\sqrt{t_2}\end{pmatrix}\begin{pmatrix}e^{i2\theta}&0\\0&1\end{pmatrix}\begin{pmatrix}\sqrt{t_1}&i\sqrt{1-t_1}\\i\sqrt{1-t_1}&\sqrt{t_1}\end{pmatrix}\begin{pmatrix}e^{i\phi}&0\\0&1\end{pmatrix},\nonumber \\
    &= \sqrt{1-\alpha}\begin{pmatrix}\sqrt{t_1t_2}e^{i2\theta}e^{i\phi} + \sqrt{(1-t_1)(1-t_2)}e^{i\phi} & i\sqrt{t_1(1-t_2)}e^{i2\theta} - i\sqrt{t_2(1-t_1)} \\ -i\sqrt{t_2(1-t_1)}e^{i2\theta}e^{i\phi} + i\sqrt{t_1(1-t_2)}e^{i\phi} & \sqrt{(1-t_1)(1-t_2)}e^{i2\theta} + \sqrt{t_1t_2}\end{pmatrix}.
\end{align}
\end{widetext}
In the Supplementary Information S1, we analyze the regimes in $\alpha_\mathrm{WG}$, $L$ where the imperfect DC splitting ratios are dominant, and vice versa.

\subsection*{Network Optimization \& Training Processes}
A QPNN is trained to perform a mapping between a set of $K$ input-output state pairs $\left|\psi_\mathrm{in}^{(i)}\right\rangle \to$ $\left|\psi_\mathrm{out}^{(i)}\right\rangle$. For the QPNN-based BSA, the training set is provided in the computational basis in Fig~\ref{fig:overview}a. Since dual-rail encoding is applied, $\left|0\right\rangle$ $\left(\left|1\right\rangle\right)$ in the computational basis is equivalent to $\left|10\right>$ $\left(\left|01\right\rangle\right)$ in the Fock basis for the two spatial modes that realize the photonic qubit. 

The unconditional infidelity of the network, $\mathcal{C}^{(\mathrm{unc})} = 1-\mathcal{F}^{(\mathrm{unc})}$ (see Eqs.~\ref{eq:Func_i},\ref{eq:Func} for $\mathcal{F}^{(\mathrm{unc})}$), is minimized to facilitate the optimization process. The variational parameters, $\left\{\boldsymbol{\phi}_i, \boldsymbol{\theta}_i\right\}$ for each layer in the network, are initialized randomly. Then, the local, gradient-free BOBYQA nonlinear optimization algorithm \cite{BOBYQA} (available from the NLopt library \cite{NLopt}) is applied until the absolute change in infidelity is less than some threshold chosen empirically based on the available computational resources. This algorithm constructs a quadratic approximation to the infidelity and thus does not require an analytical gradient. Gradient-free optimization was deemed pertinent since it is unlikely that the internal state of the network, as would be necessary for backpropagation methods, would be accessible during \emph{in situ} training \cite{Steinbrecher:19}.

In contrast to \emph{in situ} training, as described in the main text, offline training was conducted by training a QPNN with idealized components, then adding fabrication imperfections, and if necessary, adjusting the effective nonlinearity (c.f. Sec.~\ref{sec:wni}). Due to the loss and DC splitting ratio variations, such imperfections were added to an idealized solution in 50 (200) repetitions in Fig.~\ref{fig:loss} (\ref{fig:nl}), matching the number of \emph{in situ} trials conducted. From these results, an \emph{in situ} trial was deemed successful if it achieved an optimized unconditional infidelity at or below the worst-case of offline training (mean minus standard deviation). Only successful optimization trials were considered for further analysis. Similarly, the loss limit is computed by adding imperfections to an idealized solution, however, losses are assumed to be completely uniform at $\alpha_\mathrm{WG}$, and the DC splitting ratios are all $50:50$.

All simulations were conducted on the Frontenac Platform computing cluster offered by the Centre for Advanced Computing at Queen's University. The accompanying code was written in Python (version 3.10.2) using Numpy (version 1.22.2) and NLopt (version 2.6.1). Cython (version 0.29.30) was used to translate performance-sensitive operations to C to improve computation runtime. In the Supplementary Information S1, we identify where computational complexity arises when constructing the system transfer function.

\subsection*{Conditional Measures}
As for the unconditional fidelity, the conditional fidelity can be found by projecting the actual output state, $\left|\psi_\mathrm{out,act}^{(i)}\right\rangle = \mathbf{S}\left|\psi_\mathrm{in}^{(i)}\right\rangle$, onto the computational basis, $\mathrm{CB}$, and finding its overlap with the ideal output $\left|\psi_\mathrm{out}^{(i)}\right\rangle$. Averaging over all $K$ input-output pairs, this is written as,
\begin{equation} \label{eq:Fcon}
    \mathcal{F}^{(\mathrm{con})} = \frac{1}{K}\sum_{i=1}^{K}\left|\left\langle \psi_{\mathrm{out}}^{(i)}\right|A^{(i)}\mathbf{S}\left|\psi_{\mathrm{in}}^{(i)}\right\rangle\right|^2,
\end{equation}
where, 
\begin{equation} \label{eq:Ai}
    A^{(i)} = \left[\sum_{\left|x\right\rangle\in\mathrm{CB}}\left|\bigl\langle x\bigr|\mathbf{S}\bigl|\psi_\mathrm{in}^{(i)}\bigr\rangle\right|^2\right]^{-\frac{1}{2}},
\end{equation}
normalizes the $i^{\mathrm{th}}$ $\mathbf{S}\left|\psi_{\mathrm{in}}^{(i)}\right\rangle$ to the computational basis. Similarly, the probability of measuring an output in the computational basis is
\begin{equation} \label{eq:Pcb}
    \mathcal{P}^{(\mathrm{cb})} = \frac{1}{K}\sum_{i=1}^K \sum_{\left|x\right\rangle\in\mathrm{CB}}\left|\bigl\langle x\bigr|\mathbf{S}\bigl|\psi_\mathrm{in}^{(i)}\bigr\rangle\right|^2.
\end{equation}
The $i^\text{th}$ terms of Eqs.~\ref{eq:Fcon}, \ref{eq:Pcb} can be multiplied to yield Eq.~\ref{eq:Func_i}, which follows simply from the fact that the $i^\text{th}$ term of Eq.~\ref{eq:Pcb} can be expressed as $\left(A^{(i)}\right)^{-2}$.

\section{Acknowledgements} \label{sec:ack}
This research is supported by the Vector Scholarship in Artificial Intelligence, provided through the Vector Institute. The authors thank N.R.H. Pedersen for his insights into linear meshes, and gratefully acknowledge support by the Natural Sciences and Engineering Research Council of Canada (NSERC), the Canadian Foundation for Innovation (CFI), and Queen's University.

\section{Author Contributions} \label{sec:auth}
N.R. and J.C. conceived the project, which they developed along with J.E. J.E. was responsible for designing and performing all simulations and analysis, with supervision from B.S. and N.R. All authors discussed the results and shared in the writing and editing responsibilities for the manuscript.

\section{Additional Information} \label{sec:add}
\noindent
\textbf{Supplementary Information} accompanies the paper.

\noindent
\textbf{Competing Interests:} The authors declare no competing interests.

\bibliography{Refs_manuscript.bib}

\end{document}


\renewcommand{\theequation}{S\arabic{equation}}
\renewcommand{\thefigure}{S\arabic{figure}}
\renewcommand{\bibnumfmt}[1]{[S#1]}
\renewcommand{\citenumfont}[1]{S#1}
\renewcommand{\thetable}{S\arabic{table}}
\setcounter{section}{0}
\renewcommand{\thesection}{S\arabic{section}}

\title{Supplementary Information for ``Realistic quantum photonic neural networks''}

\author{Jacob Ewaniuk}
\email{jacob.ewaniuk@queensu.ca}
\affiliation{Department of Physics, Engineering Physics \& Astronomy, 64 Bader Lane, Queen's University, Kingston, Ontario, Canada K7L 3N6}

\author{Jacques Carolan}
\affiliation{Wolfson Institute for Biomedical Research, University College London, London WC1E 6BT, UK}

\author{Bhavin J. Shastri}
\affiliation{Department of Physics, Engineering Physics \& Astronomy, 64 Bader Lane, Queen's University, Kingston, Ontario, Canada K7L 3N6}
\affiliation{Vector Institute, Toronto, Ontario, Canada, M5G 1M1}

\author{Nir Rotenberg}
\affiliation{Department of Physics, Engineering Physics \& Astronomy, 64 Bader Lane, Queen's University, Kingston, Ontario, Canada K7L 3N6}

\date{\today}

\maketitle

\section{Constructing the System Transfer Function for a Realistic Quantum Photonic Neural Network} \label{sec:transfer}
The quantum photonic neural network (QPNN) architecture, as illustrated in Fig. 1a of the main manuscript, can be described via operation in the discrete $N = {n+m-1\choose n}$-dimensional Fock basis as generated by the $m$ spatial modes of $n$ photons \cite{Steinbrecher:19}. Photonic qubits are dual-rail encoded such that $n = 2$ photons and $m = 4$ modes are required to establish the two-qubit inputs of the QPNN-based Bell-state analyzer (BSA). Explicitly, the set of two-photon Fock basis states is expressed as,
\begin{equation} \label{eq:fock}
    \left\{\left|2000\right\rangle, \left|1100\right\rangle, \left|1010\right\rangle, \left|1001\right\rangle, \left|0200\right\rangle, \left|0110\right\rangle, \left|0101\right\rangle, \left|0020\right\rangle, \left|0011\right\rangle, \left|0002\right\rangle\right\},
\end{equation}
where the subset,
\begin{equation} \label{eq:cb}
    \left\{\left|1010\right\rangle, \left|1001\right\rangle, \left|0110\right\rangle, \left|0101\right\rangle\right\},
\end{equation}
represents the computational basis such that $\left|10\right\rangle = \left|0\right\rangle_\mathrm{logic}$, $\left|01\right\rangle = \left|1\right\rangle_\mathrm{logic}$.

The $i^\text{th}$ layer of a QPNN consists of a linear unitary transformation $\tilde{\mathbf{U}}(\boldsymbol{\phi}_i,\boldsymbol{\theta}_i)$ that is modelled according to the encoding scheme of Ref.~\cite{Clements:16}. This scheme involves a rectangular mesh of Mach-Zehnder interferometers (MZIs), each corresponding to a unitary transformation $\mathbf{T}_{p,q}(\phi, \theta)$ on modes $p$, $q$, and a set of single-mode phase shifts at the output of the mesh as described by the diagonal unitary $\mathbf{D}$. In terms of these transformations, $\tilde{\mathbf{U}}(\boldsymbol{\phi}_i,\boldsymbol{\theta}_i)$ is defined as,
\begin{equation} \label{eq:clements}
    \tilde{\mathbf{U}}(\boldsymbol{\phi}_i,\boldsymbol{\theta}_i) = \mathbf{D}\prod_{(p,q)\in R}\mathbf{T}_{p,q}(\phi, \theta),
\end{equation}
where $R$ is a sequence of the $m(m-1)/2$ two-mode transformations, and $\phi$, $\theta$ are elements of the corresponding vectors $\boldsymbol{\phi}_i,\boldsymbol{\theta}_i$ that are selected according to the sequence \cite{Clements:16}. This $m$ x $m$ unitary transformation correctly describes the photonic circuit at layer $i$, however, state propagation is resolved into the $N$-dimensional basis of Eq. \ref{eq:fock} in the simulations, where $N \neq m : n > 1$. Thus, a multi-photon unitary transformation, $\mathbf{U} = \Phi(\tilde{\mathbf{U}})$, is applied as defined by Eq. 6 of Ref.~\cite{Aaronson:10}. This transformation requires the calculation of $N^2$ permanents of $n \times n$ matrices, making it by far the most computationally demanding process in the simulations \cite{Steinbrecher:19}. Between consecutive QPNN layers, single-site Kerr nonlinearities are assumed. These components generate an $N$ x $N$ diagonal unitary of the form,
\begin{equation} \label{eq:kerr}
    \boldsymbol{\Sigma}(\varphi) = \sum_n \exp{\left[in(n-1)^\frac{\varphi}{2}\right]}\left|n\right\rangle\left\langle n\right|,
\end{equation}
where $\varphi$ is the effective nonlinear phase shift \cite{Steinbrecher:19}. By left-multiplying each transformation from the input to the output of the network architecture (c.f. Fig. 1a of the main manuscript), the system transfer function is defined as,
\begin{equation} \label{eq:system}
    \mathbf{S} = \mathbf{U}(\boldsymbol{\phi}_L, \boldsymbol{\theta}_L) \cdot \prod_{i=1}^{L-1} \boldsymbol{\Sigma}(\varphi) \cdot \mathbf{U}(\boldsymbol{\phi}_i, \boldsymbol{\theta}_i),
\end{equation}
for a QPNN of $L$ layers.

\subsection*{Additional Details on the Fabrication Imperfection Model} \label{sec:transfer-fab}
\begin{figure}[htb]
    \centering
    \includegraphics[scale=1.0]{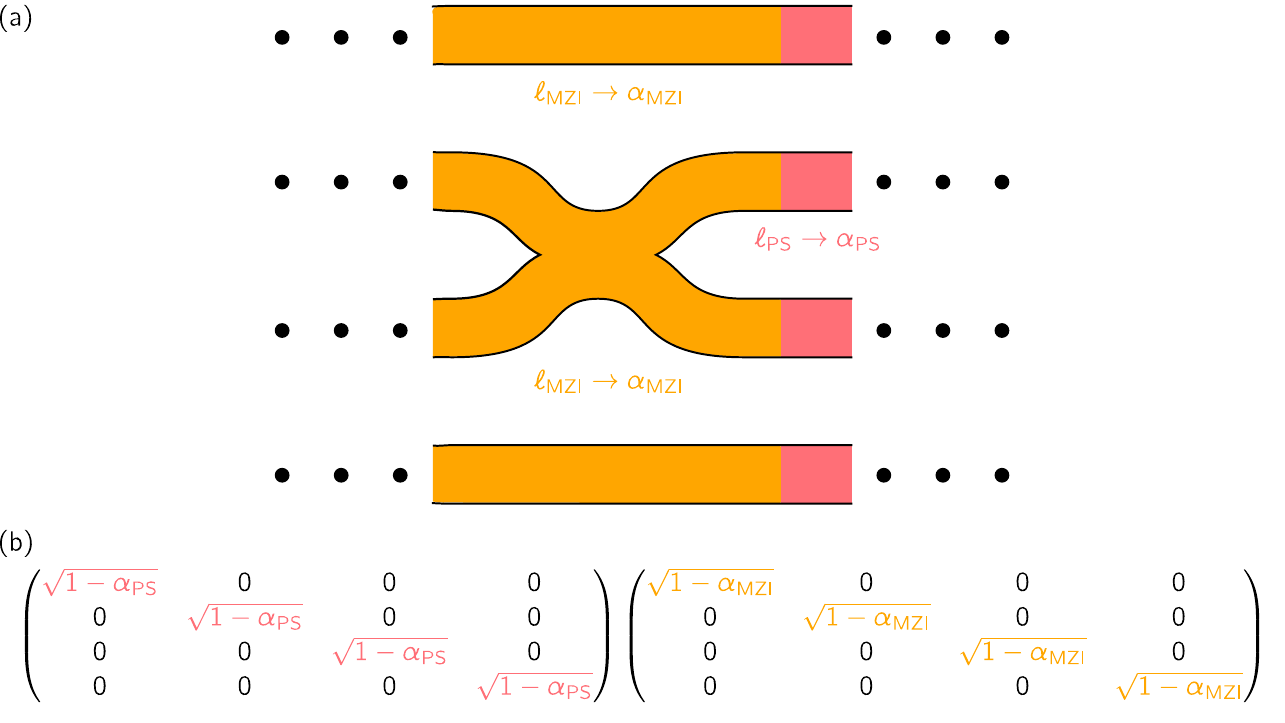}
    \caption{(a) Diagrammatic representation of the characteristic lengths identified for the propagation loss model of a realistic QPNN. Each component of length $\ell$ contributes the probability $\alpha$ of losing a photon. The flat sections adjacent to a MZI must share its length for the circuit to be balanced. (b) Matrices, with color-coded elements, constructed to apply losses in the simulations for the components shown in (a).}
    \label{fig:lengths}
\end{figure}

Waveguide propagation losses are proportional to waveguide length \cite{Sala:16}. Thus, the characteristic lengths of the components in the QPNN architecture were identified as displayed in Fig.~\ref{fig:lengths}a. 
The single-site Kerr nonlinearities were assumed to be much shorter in length than the elements of a MZI mesh such that their losses were deemed negligible. To determine the characteristic lengths, we considered components fabricated for 1550 nm operation in silicon-on-insulator (SOI) that do not contribute excess loss. Specifically, the chosen broadband directional coupler (DC), with 5.08\% splitting ratio variations as discussed in the main text, has a waveguide length of 93.342 $\mu$m when two 10 $\mu$m bends (input and output) are added to the design of Ref.~\cite{Lu:15}. The phase shifters were modelled after the 50 $\mu$m-long thermo-optic indium tin oxide design of Ref.~\cite{Parra:20}, where $< 0.01$ dB insertion losses were reported (negligible excess loss). Since each MZI has two phase shifters and two DCs, the characteristic lengths are defined as $\ell_\mathrm{MZI} = 286.684$ $\mu$m and $\ell_\mathrm{PS} = 50$ $\mu$m respectively. These lengths are in agreement with those reported as typical in Ref.~\cite{Sala:16} for SOI thermo-optic phase shifters and DCs when bend radii are chosen such that excess losses are negligible.

When losses are included $\left(\alpha_\mathrm{WG} \neq 0\right)$ in the QPNN simulations, the MZI transformation (c.f. Eq. 10 in the main manuscript) becomes non-unitary, and each linear transformation $\mathbf{U}(\boldsymbol{\phi}_i, \boldsymbol{\theta}_i)$ follows. When applied to a normalized state resolved in the Fock basis, the resultant state lacks normalization. At the output of the network, the `missing' coefficient that would yield a normalized state can be regarded as attached to some state $\left|\epsilon_\mathrm{loss}\right\rangle$, appended to the basis of Eq.~\ref{eq:fock}, that accounts for all potential output states where $n < 2$ across the four spatial modes of interest. For our purposes, it is sufficient that the loss model accurately reduces the probability of measuring a state in the ideal Fock basis (Eq.~\ref{eq:fock}), and thus that of measuring the target output state, without fully describing $\left|\epsilon_\mathrm{loss}\right\rangle$.

\begin{figure}[htb!]
\centering
\includegraphics[scale=1.0]{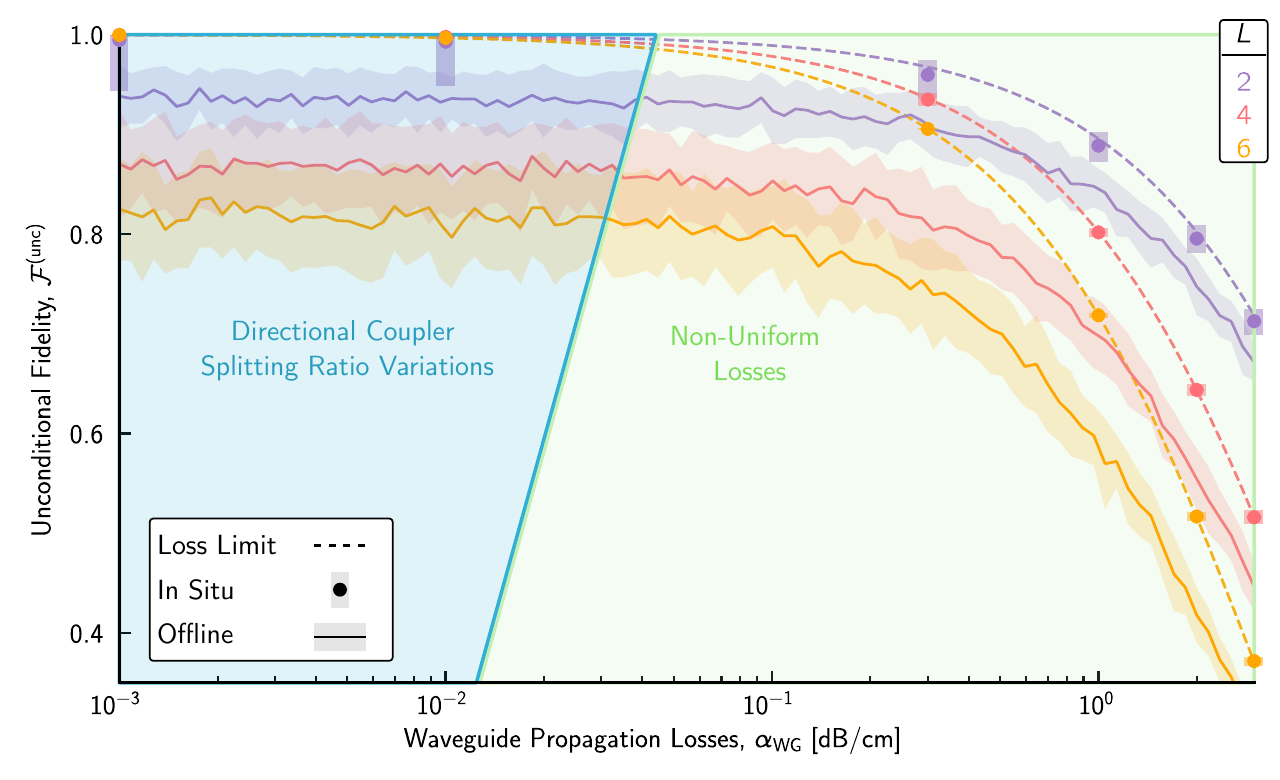}
\caption{Identifying dominant QPNN fabrication imperfections. The unconditional fidelity is plotted for QPNNs of 2, 4, and 6 layers, with ideal nonlinearities $\left(\varphi = \pi\right)$, as a function of losses. Both \emph{in situ} and offline-trained QPNNs feature loss and splitting ratio variations that match the model applied in the main manuscript. The shaded blue and green regions were added as a visual aid to show the domains in $\alpha_\mathrm{WG}$ where DC splitting ratio variations and non-uniform losses dominate, respectively.}
\label{fig:dcvsloss}
\end{figure}

The DC splitting ratio variations are independent of propagation losses. Thus, there exists a regime where the losses are small enough such that the splitting ratio variations become the dominant imperfection. This is evident in the main manuscript (Fig. 2g) when considering the increase to the unconditional fidelity $\mathcal{F}^{(\mathrm{unc})}$ from $L = 2$ to $L = 4$ for $\alpha_\mathrm{WG} \leq 0.01$ dB/cm. It is more explicitly shown in Fig.~\ref{fig:dcvsloss}, where $\mathcal{F}^{(\mathrm{unc})}$ is plotted for QPNNs of 2 to 6 layers.
The shaded blue region highlights the regime where the DC splitting ratio variations are dominant. In this regime, an increase to the size of the QPNN does not yield a decrease in $\mathcal{F}^{(\mathrm{unc})}$ for those trained $\emph{in situ}$. In fact, the unconditional fidelity often increases due to the additional parameters that can be used to learn how to optimize around the fabrication imperfections. This further explains why $\mathcal{F}^{(\mathrm{unc})}$ decreases for QPNNs trained offline, regardless of $\alpha_\mathrm{WG}$. Conversely, in the loss-dominant regime (shaded green region), additional layers only hinder the performance of the QPNN, given that it has ideal nonlinearities $\left(\varphi = \pi\right)$ in this case, regardless of \emph{in situ} or offline training.

\section{Optimizing the Conditional Fidelity} \label{sec:con}
\begin{figure}[ht!]
\centering
\includegraphics[width=\textwidth]{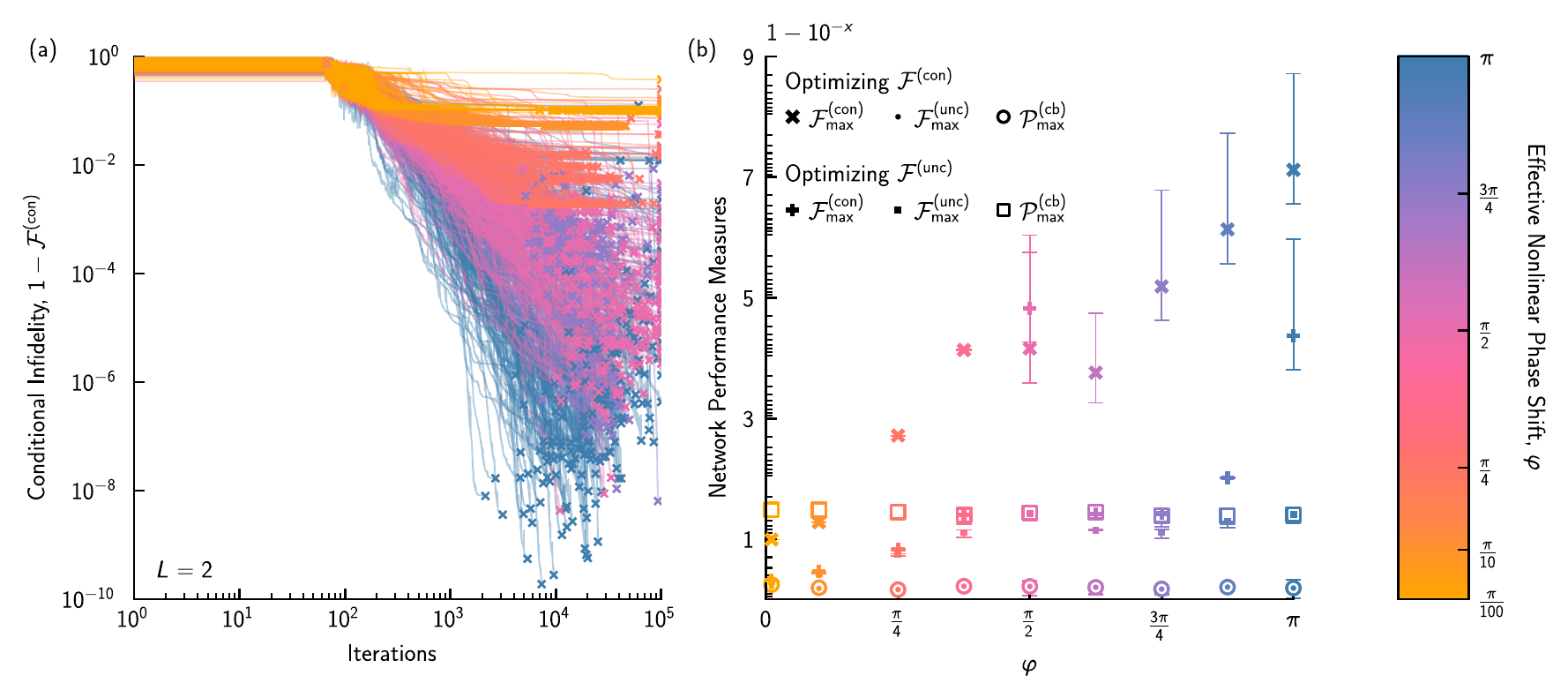}
\caption{Training realistic 2-layer QPNNs with state-of-the-art losses $\left(\alpha_\mathrm{WG} = 0.03\text{ dB}/\text{cm}\right)$ and weak nonlinearities to optimize conditional fidelity $\mathcal{F}^{(\mathrm{con})}$. (a) Training traces for QPNNs, with effective nonlinear phase shifts as labelled in the colorbar, when the conditional infidelity is minimized. (b) Conditional fidelities $\mathcal{F}^{(\mathrm{con})}$, unconditional fidelities $\mathcal{F}^{(\mathrm{unc})}$, and computational basis probabilities $\mathcal{P}^{(\mathrm{cb})}$ for QPNNs of varying effective nonlinear phase shifts $\varphi$ when trained to optimize $\mathcal{F}^{(\mathrm{con})}$ and $\mathcal{F}^{(\mathrm{unc})}$ respectively. Error bars show the 95\% confidence intervals of a beta distribution (see Sec.~\ref{sec:stat} for more details).}
\label{fig:con}
\end{figure}

In Fig.~\ref{fig:con}a, we provide training traces for realistic 2-layer QPNNs with weak nonlinearities $\left(\varphi \lesssim \pi\right)$, assuming state-of-the-art losses $\left(\alpha_\mathrm{WG} = 0.03\text{ dB}/\text{cm}\right)$, when minimizing the conditional infidelity, $1 - \mathcal{F}^{(\mathrm{con})}$, rather than the unconditional. 
All training procedures were chosen to match those explained in the main manuscript. Thus, the absence of plateaus for $\varphi \geq \pi / 2$ may be attributable to sub-optimal training parameters. The QPNN performance measures in this case are compared to those achieved when minimizing the unconditional infidelity, as in the main manuscript, in Fig.~\ref{fig:con}b. In general, when a QPNN is trained to minimize its conditional infidelity, it is able to achieve multiple order-of-magnitude improvements at the expense of significantly reduced operational efficiency. Taking $\varphi = 3\pi / 4$ as an example, the conditional fidelity increases from 0.851 (0.846) to 0.998 (0.998) while the probability of yielding a logical output decreases from 0.964 (0.957) to 0.312 (0.312), correlating to a decrease in unconditional fidelity from 0.820 (0.809) to 0.311 (0.311), where the bracketed results show the lower bounds of 95\% confidence intervals. Overall, if an application prioritizes $\mathcal{F}^{(\mathrm{con})}$, realistic QPNNs have the versatility to meet this desire.

\section{Statistical Analysis} \label{sec:stat}
\begin{figure}[b]
\centering
\includegraphics[width=\textwidth]{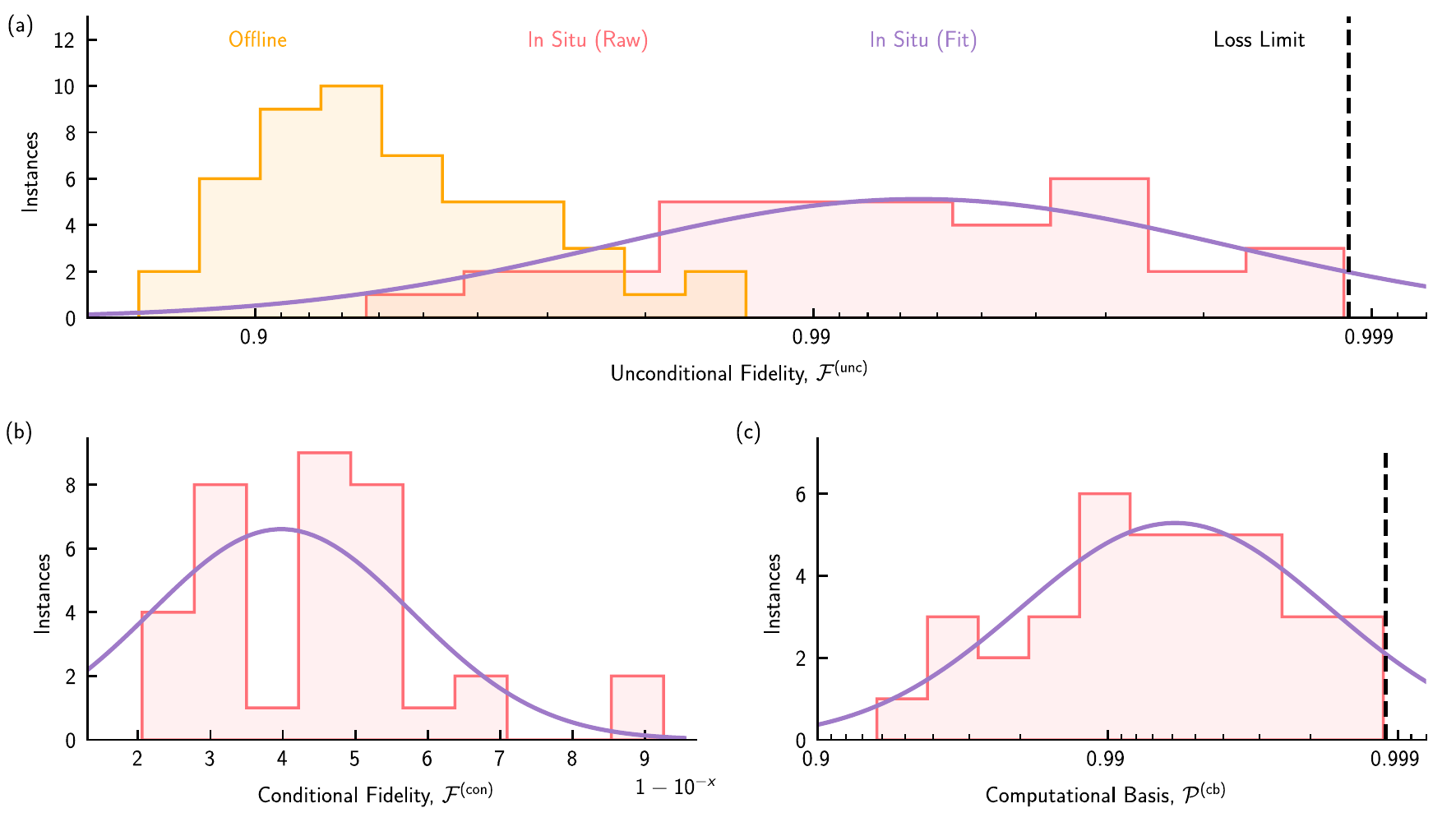}
\caption{Data analysis for successfully-trained 2-layer realistic QPNNs of ideal nonlinearities $\left(\varphi = \pi\right)$ and 0.01 dB/cm losses. (a) Histograms of the unconditional fidelities $\mathcal{F}^{(\mathrm{unc})}$ for QPNNs trained offline (orange) and \emph{in situ} (pink). The \emph{in situ} data was fit with a logarithmic normal distribution (purple line), and the loss limit (dashed black line, also in (c)) is shown for comparison. The (b) conditional fidelities $\mathcal{F}^{(\mathrm{con})}$ and (c) computational basis probabilities $\mathcal{P}^{(\mathrm{cb})}$ for QPNNs trained \emph{in situ} are given as histograms (pink) with respective logarithmic normal distribution fits (purple lines).}
\label{fig:hist_loss}
\end{figure}
\begin{figure}[b]
\centering
\includegraphics[width=\textwidth]{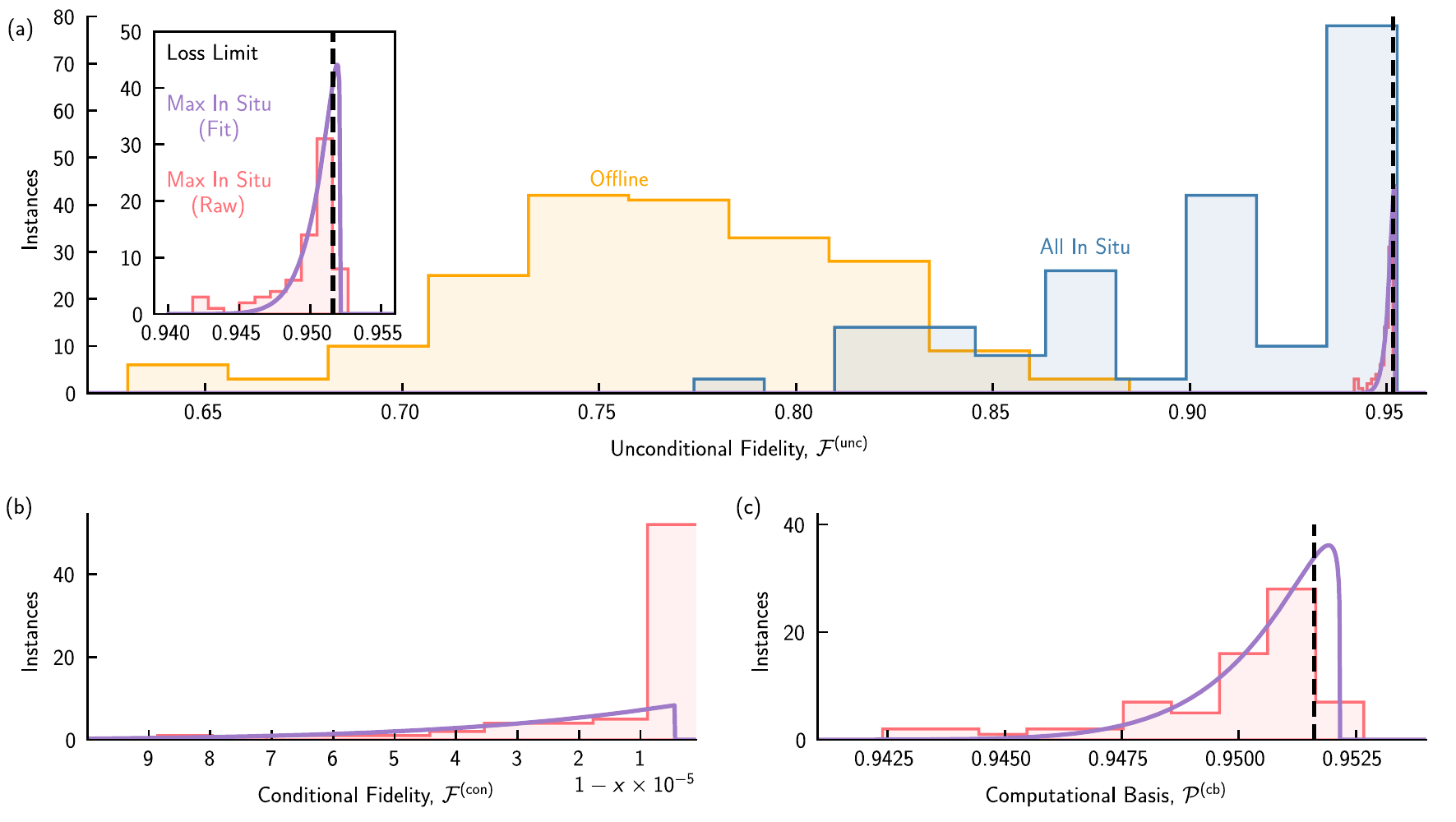}
\caption{Data analysis for successfully-trained 3-layer realistic QPNNs of effective nonlinearity $\varphi = 3\pi / 4$ and 0.3 dB/cm losses. (a) Histograms of the unconditional fidelities $\mathcal{F}^{(\mathrm{unc})}$ for QPNNs trained offline (orange) and \emph{in situ} (blue). A separate histogram (pink) is displayed that only contains the \emph{in situ} results for QPNNs trained to the minimum plateau in unconditional infidelity (c.f. Fig.~\ref{fig:train_nl}a), and the inset provides an expanded view of it. This histogram was fit with a beta distribution (purple line), and the loss limit (dashed black line, also in (c)) is shown for comparison. The (b) conditional fidelities $\mathcal{F}^{(\mathrm{con})}$ and (c) computational basis probabilities $\mathcal{P}^{(\mathrm{cb})}$ for QPNNs trained \emph{in situ} are given as histograms (pink) with respective beta distribution fits (purple lines).}
\label{fig:hist_nl}
\end{figure}

In all realistic QPNN simulations, there are trials where the optimization process reaches a local minimum and converges to a solution that can be deemed a failure. To eliminate these trials from the analysis of the network performance measures, successful training thresholds were computed for each combination of network size ($L$), losses $\left(\alpha_\mathrm{WG}\right)$, and effective nonlinearity ($\varphi$), examples of which are displayed in Figs. 2a-c of the main manuscript and Figs.~\ref{fig:train_nl}a-c. A given trial is deemed to have successfully trained the QPNN if the optimized unconditional fidelity $\mathcal{F}^{(\mathrm{unc})}$ is greater than the lower bound (mean minus standard deviation) of the corresponding offline training result (c.f. lower shaded grey regions in Figs. 2d-f, 3a-c of the main manuscript). When considering realistic QPNNs with weak nonlinearities (c.f. Fig. 3 of the main manuscript, Figs.~\ref{fig:con}, \ref{fig:fredkin}, \ref{fig:nl_ghz}), trials were often separated into a set of plateaus in unconditional infidelity, as visually evident in Fig.~\ref{fig:train_nl}a. In these cases, the trials reaching the minimum plateau were further isolated for analysis, denoting the corresponding performance measures with the subscript `max' (minimum infidelity corresponds to maximum fidelity).  Once trials were appropriately isolated, the raw data was fit to a matching probability distribution to compute the results plotted in Figs. 2d-i, 3a-f of the main manuscript.

\subsection*{Fitting the Propagation Loss Results} \label{sec:stat-loss}
In Fig.~\ref{fig:hist_loss}, we provide the raw data for the successfully-trained 2-layer realistic QPNNs of ideal nonlinearities $\left(\varphi = \pi\right)$ and 0.01 dB/cm losses, as an example. The offline training data is compared with the \emph{in situ} in Fig.~\ref{fig:hist_loss}a, showing the general increase in unconditional fidelity that is achieved as the network learns to account for fabrication imperfections, thus approaching the loss limit. The \emph{in situ} unconditional fidelity (Fig.~\ref{fig:hist_loss}a), conditional fidelity (Fig.~\ref{fig:hist_loss}b), and computational basis probability (Fig.~\ref{fig:hist_loss}c) data were respectively fit with logarithmic normal distributions using SciPy (version 1.8.1). The mean (95\% confidence intervals) of each of these distributions provide the points (error bars or shaded regions) plotted in Fig. 2 of the main manuscript.

\subsection*{Fitting the Weak Nonlinearity Results} \label{sec:stat-nl}
In Fig.~\ref{fig:hist_nl}, we provide the raw data for the successfully-trained 3-layer realistic QPNNs of effective nonlinearity $\varphi = 3\pi/4$ and state-of-the-art 0.3 dB/cm losses, as an example. The offline training data is compared with the \emph{in situ} in Fig.~\ref{fig:hist_nl}a, showing the general increase in unconditional fidelity that is achieved as the network learns to account for weak nonlinearities and fabrication imperfections, thus approaching the loss limit. There are clearly separated plateaus in unconditional fidelity (separate peaks in the blue all \emph{in situ} histogram), as can also be viewed in terms of infidelity Fig.~\ref{fig:train_nl}b. Only trials within the maximum unconditional fidelity plateau were further analyzed by fitting respective beta distributions to the unconditional fidelities (inset to Fig.~\ref{fig:hist_nl}a), conditional fidelities (Fig.~\ref{fig:hist_loss}b), and computational basis probabilities (Fig.~\ref{fig:hist_loss}c) using SciPy (version 1.8.1). The mean (95\% confidence intervals) of each of these distributions provide the points (error bars or shaded regions) plotted in Fig. 3 of the main manuscript.

\section{Training Networks with Weak Nonlinearities} \label{sec:wnl}
\begin{figure}[htb!]
\centering
\includegraphics[width=\textwidth]{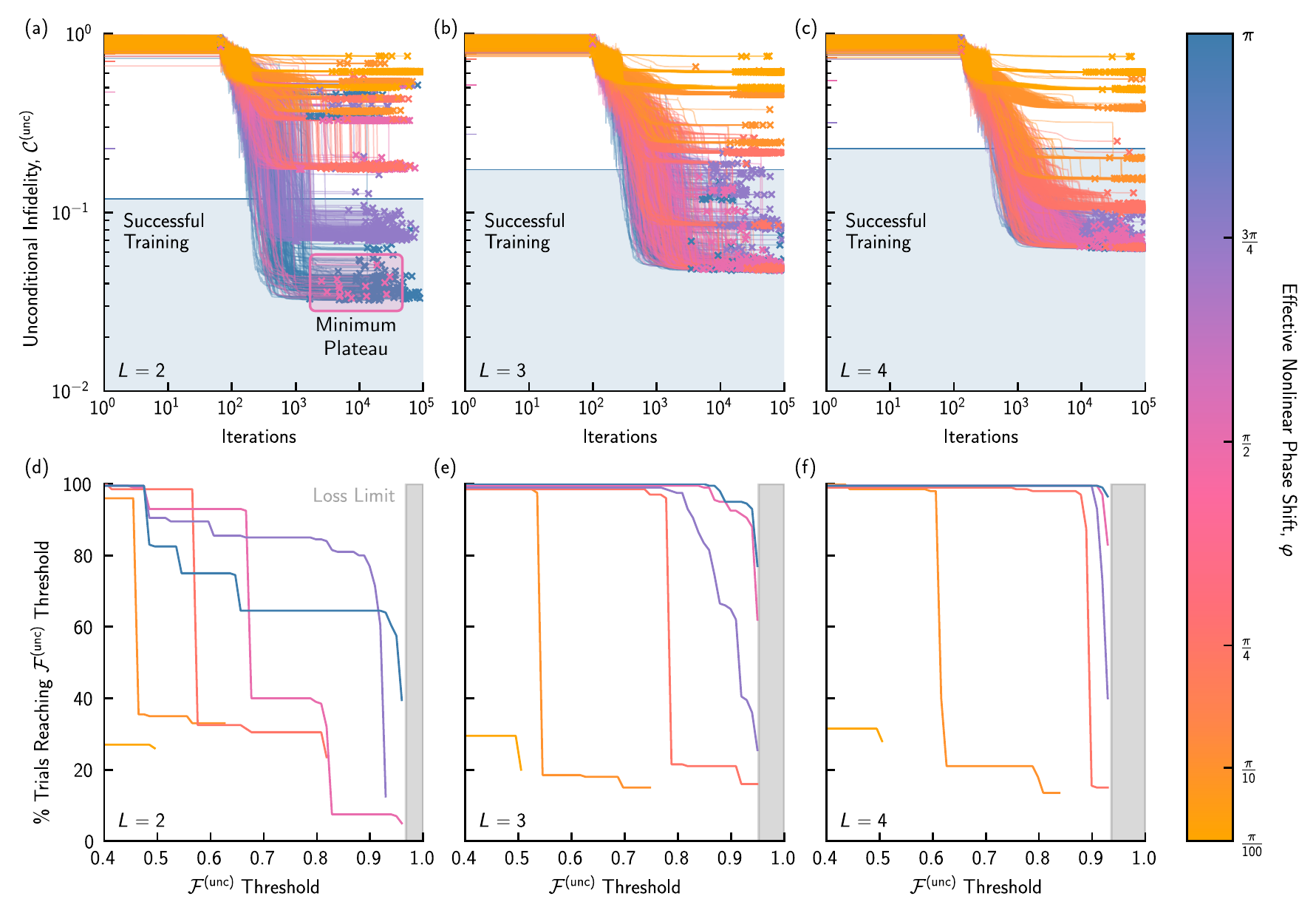}
\caption{Training data for realistic QPNNs with state-of-the-art 0.3 dB/cm losses and weak nonlinearities $\left(\varphi \lesssim \pi\right)$. The unconditional infidelity $\mathcal{C}^{(\mathrm{unc})}$ of (a) 2, (b) 3, and (c) 4-layer networks are shown as a function of the training iteration for networks with effective nonlinearities as labelled in the colorbar. Colored ticks denote the thresholds computed to determine whether the QPNN was successfully-trained in a given trial, out of 200 trials each, with the shaded blue region displaying an example for $\varphi = \pi$. In (a), the minimum plateau in $\mathcal{C}^{(\mathrm{unc})}$ for QPNNs of $\varphi = \pi/2$ is displayed as a visual aid. The trainability of (d) 2, (e) 3, and (f) 4-layer networks is examined by plotting the percentage of trials that achieve an unconditional fidelity threshold, over a domain of them. The shaded grey region depicts the $\mathcal{F}^{(\mathrm{unc})}$ that cannot be achieved due to the loss limit.}
\label{fig:train_nl}
\end{figure}

In Figs.~\ref{fig:train_nl}a-c, we provide exemplary training traces for realistic QPNNs of 2 to 4 layers with state-of-the-art losses $\left(\alpha_\mathrm{WG} = 0.3\text{ dB}/\text{cm}\right)$ and weak Kerr nonlinearities $\left(\varphi \lesssim \pi\right)$.
For each effective nonlinear phase shift $\varphi$, 200 optimization trials were conducted. The trainability of these QPNNs is examined in Figs.~\ref{fig:train_nl}d-f, where the percentage of trials that reach a unconditional fidelity threshold is plotted over a domain of $\mathcal{F}^{(\mathrm{unc})}$. Here, the difficulty of training QPNNs, with $\varphi < \pi$, to loss-limited performance is evident. QPNNs of 2 layers (Fig.~\ref{fig:train_nl}d) can reach optimal performance with $\varphi = \pi / 2$ in addition to $\pi$, however, the desired unconditional fidelity is only reached in 5.0\% of trials for the former, compared to 39.5\% of trials for the latter. By increasing the network size, the trainability improves in general, for all $\varphi$, at the cost of a decreased loss limit. 

\subsection*{Comparing Realistic QPNNs with an Ideal Fredkin Gate-based BSA} \label{sec:fredkin}
\begin{figure}[t]
\centering
\includegraphics[scale=1.0]{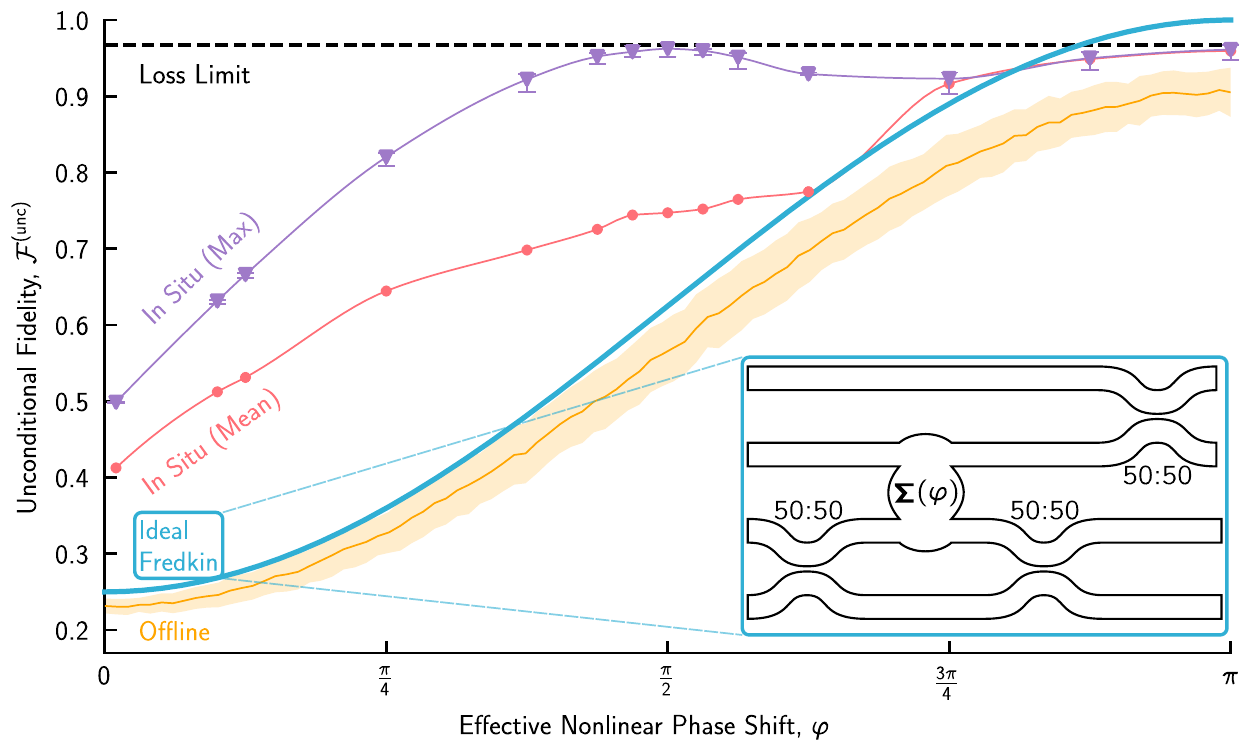}
\caption{Comparison between the respective unconditional fidelities of perfectly-fabricated quantum-optical Fredkin gate-based BSAs and realistic 2-layer QPNN-based BSAs with state-of-the-art 0.3 dB/cm losses, as a function of effective nonlinear phase shift $\varphi$. The inset shows a photonic circuit diagram of the Fredkin gate-based BSA. Results are shown for QPNNs trained offline (orange line shows mean, shaded region shows standard deviation) and \emph{in situ}, both as the mean of all trials (pink circles) and the results at the maximum unconditional fidelity plateau (purple triangles with error bars showing 95\% confidence intervals), which are all constrained by the loss limit (dashed black line). Connecting lines for the \emph{in situ} results serve as a visual aid.}
\label{fig:fredkin}
\end{figure}

A BSA can be constructed from a controlled-NOT (CNOT) gate followed by a Hadamard gate applied to the control qubit \cite{Nielsen:11}. With dual-rail encoded photonic qubits, the CNOT gate can be realized from a quantum-optical Fredkin gate \cite{Milburn:89} which uses two 50:50 DCs (equivalent to Hadamard gates) and a nonlinear Kerr medium that connects the two photonic qubits as displayed in the inset to Fig.~\ref{fig:fredkin}.
However, it is only deterministic for an effective nonlinear phase shift $\varphi = \pi$ \cite{Gerry:04}. Altogether, this circuit can be described by the unitary transformation,
\begin{align} \label{eq:fredkinU}
    \mathbf{U}_\mathrm{BSA} &= \left(\mathbf{H}\otimes \mathbf{I}\right)\cdot\left(\mathbf{I}\otimes \mathbf{H}\right)\cdot\boldsymbol{\Sigma}(\varphi)\cdot\left(\mathbf{I}\otimes \mathbf{H}\right),\nonumber \\
    &= \frac{1}{\sqrt{2}}\begin{pmatrix}1 & 0 & 1 & 0 \\ 0 & 1 & 0 & 1 \\ 1 & 0 & -1 & 0 \\ 0 & 1 & 0 & -1\end{pmatrix}\frac{1}{\sqrt{2}}\begin{pmatrix}1 & 1 & 0 & 0 \\ 1 & -1 & 0 & 0 \\ 0 & 0 & 1 & 1 \\ 0 & 0 & 1 & -1\end{pmatrix}\begin{pmatrix} 1 & 0 & 0 & 0 \\ 0 & 1 & 0 & 0 \\ 0 & 0 & 1 & 0 \\ 0 & 0 & 0 & e^{i\varphi}\end{pmatrix}\frac{1}{\sqrt{2}}\begin{pmatrix}1 & 1 & 0 & 0 \\ 1 & -1 & 0 & 0 \\ 0 & 0 & 1 & 1 \\ 0 & 0 & 1 & -1\end{pmatrix},
\end{align}
in the computational basis of the two photonic qubits, where $\mathbf{H}$ represents the Hadamard gate, $\mathbf{I}$ is the identity, and $\boldsymbol{\Sigma}(\varphi)$ describes the transformation of effective nonlinearity $\varphi$ conducted by the Kerr medium. With this BSA transformation, the unconditional fidelity of the Fredkin gate-based BSA is given by,
\begin{equation} \label{eq:fredkinF}
    \mathcal{F}^{(\mathrm{unc})} = \frac{1}{K}\sum_{i=1}^K \left|\left\langle\psi_\mathrm{out}^{(\mathrm{i})}\right|\mathbf{U}_\mathrm{BSA}\left|\psi_\mathrm{in}^{(\mathrm{i})}\right\rangle\right|^2,
\end{equation}
for the set of $K$ input-output state pairs $\left|\psi_\mathrm{in}^{(\mathrm{i})}\right\rangle\to\left|\psi_\mathrm{out}^{(\mathrm{i})}\right\rangle$ that define the BSA operation. In this case, the unconditional fidelity is the same as the conditional since the model has considered ideal operation (i.e. no losses, DC splitting ratio variations) outside of a potentially sub-optimal $\varphi$. In Fig.~\ref{fig:fredkin}, we compare the performance of a perfectly-fabricated Fredkin gate-based BSA to the unconditional fidelities of realistic 2-layer QPNN-based BSAs with weak nonlinearities (c.f. Fig. 3a of the main manuscript), as a function of the effective nonlinear phase shift $\varphi$. The fidelity of the Fredkin gate-based BSA closely resembles the upper bound of QPNNs trained offline, yet is not hindered by the loss limit due to the absence of losses in the model described for it. Realistic QPNNs trained \emph{in situ} show improved performance for $\varphi \leq 3\pi/4$, reaffirming the learning capabilities demonstrated by the QPNN, even with weak nonlinearities. This comparison also shows why the non-monotonic relationship between $\mathcal{F}^{(\mathrm{unc})}$ and $\varphi$ for \emph{in situ}-trained QPNNs, as described in the main manuscript, was unexpected.

\section{Generating Greenberger-Horne-Zeilinger States} \label{sec:ghz}
\begin{figure}[ht!]
\centering
\includegraphics[width=\textwidth]{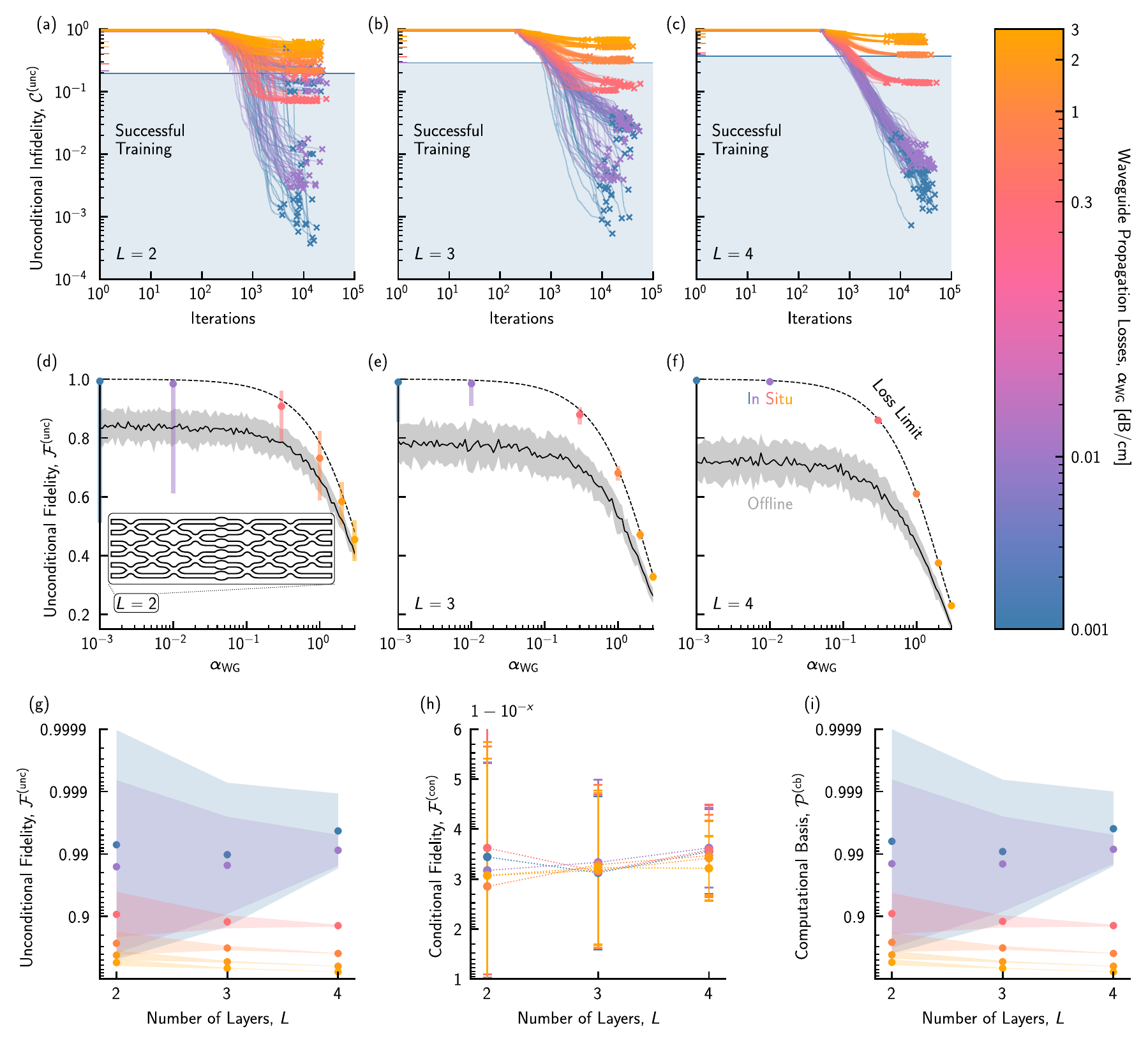}
\caption{Performance of a QPNN-based GHZ generator suffering from fabrication imperfections. The unconditional infidelity $\mathcal{C}^{(\mathrm{unc})}$ of (a) 2, (b) 3, and (c) 4-layer networks are shown as a function of the training iteration for increasingly lossy networks. In each pane, the results of 50 optimization trials are displayed, with clear plateaus visible in $\mathcal{C}^{(\mathrm{unc})}$ that increase with the losses. In each case, only trials that result in infidelity at or below those achieved by offline training (colored ticks in (a)-(c), shaded regions in (d)-(f)) are considered successful (shaded blue region shows an example for 0.001 dB/cm). The unconditional fidelity $\mathcal{F}^{(\mathrm{unc})}$ of (d) 2, (e) 3, and (f) 4-layer networks are plotted with respect to the average losses $\alpha_\mathrm{WG}$, with colored symbols (shaded regions) corresponding to the mean ($95\%$ confidence interval) of a logarithmic normal distribution fitted to the successful trials of (a)-(c). These points are seen to lie on the (dashed) loss limit curve, where the performance of the network is only limited by uniform photon loss (assumes perfect DCs), in contrast to networks that are trained offline (solid black curves and shaded grey regions), demonstrating the ability of QPNNs to learn to overcome imperfections. (g) Unconditional fidelity $\mathcal{F}^{(\mathrm{unc})}$, (h) conditional fidelity $\mathcal{F}^{(\mathrm{con})}$, and (i) computational basis probability, $\mathcal{P}^{(\mathrm{cb})}$, as a function of $L$ for the QPNNs trained \emph{in situ}, where the mean (symbols) and 95\% confidence intervals (shaded regions in (g), (i), error bars in (h)) are determined via the same method as (d)-(f).}
\label{fig:loss_ghz}
\end{figure}

\begin{figure}[ht!]
\centering
\includegraphics[width=\textwidth]{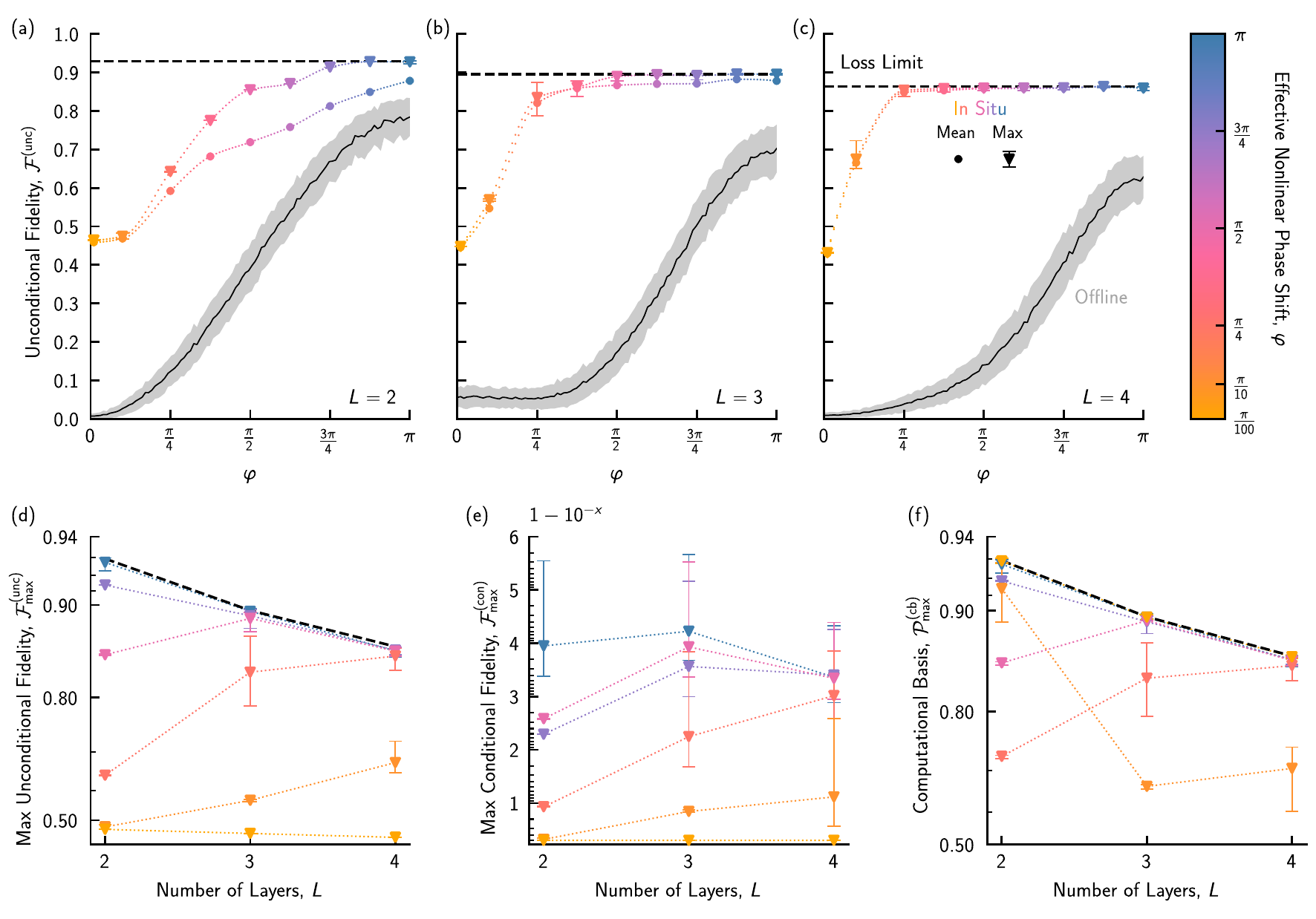}
\caption{Performance of realistic QPNN-based GHZ generators with sub-optimal $\left(\varphi\lesssim\pi\right)$ nonlinearities and state-of-the-art $\left(\alpha_\mathrm{WG} = 0.3\text{ dB}/\text{cm}\right)$ losses. The unconditional fidelity $\mathcal{F}^{(\mathrm{unc})}$ of (a) 2, (b) 3, and (c) 4-layer networks is shown with respect to the effective nonlinear phase shift $\varphi$, showing both offline (black curves, shaded grey regions) and \emph{in situ} (colored symbols) trained networks, and the loss limit (dashed line). \emph{In situ} results include the average of all successfully-trained QPNNs (circles) and the best-case, where triangles (error bars) show the mean (95\% confidence intervals) of a beta distribution fit to the maximal unconditional fidelity plateau. The (d) unconditional fidelity $\mathcal{F}^{(\mathrm{unc})}$, (e) conditional fidelity $\mathcal{F}^{(\mathrm{con})}$, and (f) computational basis probability $\mathcal{P}^{(\mathrm{cb})}$ are plotted for each $\varphi$ denoted on the colorbar, for networks of up to 4 layers. All means (triangles) and 95\% confidence intervals (error bars) were determined in the same manner as the best-case \emph{in situ} results of (a)-(c). Connecting dotted lines serve only as a visual aid.}
\label{fig:nl_ghz}
\end{figure}

Here, we present results for realistic QPNNs trained to generate Greenberger-Horne-Zeilinger (GHZ) states. The training set only uses one input-output state pair,
\begin{equation} \label{eq:ghz}
    \left|\psi_\mathrm{in}\right\rangle = \left|000\right\rangle \to \left|\psi_\mathrm{out}\right\rangle = \frac{1}{\sqrt{2}}\left(\left|000\right\rangle + \left|111\right\rangle\right),
\end{equation}
as in Ref.~\cite{Steinbrecher:19}, where the states are presented in the computational basis such that there are three photonic qubits (and thus six spatial modes, see the inset to Fig.~\ref{fig:loss_ghz}d for a QPNN-based GHZ generator diagram). 
All training procedures were chosen to match those explained in the main manuscript. Thus, the absence of plateaus for $\alpha_\mathrm{WG} \leq 0.01$ dB/cm in Figs.~\ref{fig:loss_ghz}a-c may be attributable to sub-optimal training parameters. Fig.~\ref{fig:loss_ghz} shows the results of 2 to 4-layer QPNN-based GHZ generators with perfect nonlinearities ($\varphi = \pi$) and varied losses $\alpha_\mathrm{WG}$ (c.f. Fig. 2 of the main manuscript), trained in 50 trials for each set of parameters. Networks trained \emph{in situ} approach the loss limit as discovered for the QPNN-based BSA. Similarly, the unconditional fidelity and computational basis probability may increase with additional layers for losses $\leq 0.01$ dB/cm. Fig.~\ref{fig:nl_ghz} shows the results of 2 to 4-layer QPNN-based GHZ generators with state-of-the-art losses $\left(\alpha_\mathrm{WG} = 0.3\text{ dB}/\text{cm}\right)$ and weak nonlinearities $\left(\varphi \lesssim \pi\right)$ (c.f. Fig. 3 of the main manuscript), trained in 200 trials for each set of parameters. Networks trained \emph{in situ} reach improved unconditional fidelities over offline training, and tend toward the loss limit, as discovered for QPNN-based BSAs. However, the relationship between $\mathcal{F}^{(\mathrm{unc})}$ and $\varphi$ is solely monotonic. Overall, for a given amount of losses and effective nonlinear phase shift, there similarly exists an optimal network size for QPNN-based GHZ generators.

\bibliography{Refs_supplement.bib}